\documentclass[aps,pra,twocolumn,nopacs,floatfix,longbibliography]{revtex4-1}
\usepackage{graphicx}
\usepackage{dcolumn}
\usepackage{bm}
\usepackage{multirow}
\usepackage{ulem}
\usepackage{color}
\usepackage{longtable}
\usepackage{subfigure}
\usepackage{amssymb}
\usepackage{amsmath}
\usepackage{upgreek}
\usepackage{latexsym,epsfig}
\usepackage{float}
\usepackage[figuresright]{rotating}
\usepackage{upgreek}
\usepackage{booktabs}

\begin{document}
	\title{Relativistic configuration-interaction and coupled-cluster calculations of Ir$^{17+}$ transition energies and properties for optical clock applications}
	
	\author{H. X. Liu$^{1,2}$}
	\author{Y. M. Yu$^{2,4}$} \email{Email: ymyu@iphy.ac.cn}
	\author{B. B. Suo$^{3}$}  \email{Email: bsuo@nwu.edu.cn}
    \author{Y. F. Ge$^1$}     \email{Email: yfge@ysu.edu.cn}
	\author{Y. Liu$^1$}     
	\affiliation{$^1$State Key Laboratory of Metastable Materials Science and Technology \& Hebei Key Laboratory of Microstructural Material Physics, School of Science, Yanshan University, Qinhuangdao 066004, China}
	\affiliation{$^2$Beijing National Laboratory for Condensed Matter Physics, Institute of Physics, Chinese Academy of Sciences, Beijing 100190, China}
	\affiliation{$^3$Institute of Modern Physics, Northwest University, Xi'an, Shanxi 710069, China}
	\affiliation{$^4$University of Chinese Academy of Sciences, 100049 Beijing, China}

\begin{abstract}
The transition energies and properties of the Ir$^{17+}$ ion are calculated using the Kramers-restricted configuration-interaction (KRCI) and Fock-space coupled-cluster (FSCC) methods within the Dirac-Coulomb-Gaunt Hamiltonian framework. These calculations show several forbidden optical transitions between the $4f^{13}5s$ ground state and the $4f^{14}$ and $4f^{12}5s^2$ excited states, underscoring their potential as candidates for optical clock applications. Additionally, key properties of the ground and low-lying excited states are reported, including Lande $g_J$ factors, lifetimes, electric dipole polarizabilities, electric quadrupole moments, hyperfine structure constants, relativistic sensitivities, Lorentz-invariance coefficient tensor, and isotope shifts. The excellent agreement between the results from the KRCI and FSCC methods demonstrates the robustness of the calculations and confirms the reliability of the proposed clock transitions.
\end{abstract}
	
\date{\today}
	
\maketitle
	
\section{Introduction}

Optical clocks represent one of the most significant advancements in modern metrology, utilizing the optical transition frequencies of atoms or ions to achieve unprecedented precision in time and frequency measurements, with uncertainties as low as $10^{-18}$ to $10^{-19}$. These clocks, including those based on Al$^+$ \cite{Brewer-PRL-2019}, Yb$^+$ \cite{Lange-PRL-2021,Tofful-Metrologia-2024}, Sr \cite{Aeppli2024,Takamoto-NP-2020}, Yb \cite{Kim2023,Zhang2022,Luo2020}, Ca$^+$ \cite{Zeng2023}, In$^+$ \cite{Hausser-PRL-2025}, etc. play an important role in establishing international time standards, validating fundamental physical theories, and exploring novel domains of physics. Moreover, optical clocks have become indispensable for investigating variations in fundamental constants \cite{Flambaum2007, Godun2014, Huntemann2014, Lange-PRL-2021a, Filzinger2023}, testing general relativity \cite{Chou2010, Delva2018, Takamoto-NP-2020, Bothwell-Nature-2022}, examining Lorentz invariance \cite{Megidish2019, Sanner2019, Dreissen2022}, advancing geodynamic studies \cite{McGrew2018}, enhancing time-frequency transfer \cite{Beloy2021a, Ellis2021}, supporting deep-space navigation \cite{Burt2021, Burt2024, Casini2023}, and probing the fifth force \cite{Berengut-PRL-2018, Berengut-PRR-2020, Brzeminski-PRD-2022, Rehbehn-PRL-2023, Dzuba-PRD-2024, Berengut-NRP-2025}.

Highly charged ions (HCIs) have recently emerged as a promising alternative for optical clock technology, owing to their unique electronic properties. HCIs exhibit extremely low sensitivity to external perturbations, enhancing their potential to achieve even higher precision and greater sensitivity to variations in fundamental constants compared to traditional neutral atom-based clocks. In recent years, several HCI candidates have been proposed as optical frequency standards, demonstrating their capability to achieve ultra-high precision. Notable reviews \cite{Safronova2018, Kozlov2018, Yu-FP-2023} and recent theoretical studies \cite{Berengut2011, Bekker2019, Yudin2014, Safronova2014a, Dzuba2015a, Dzuba2015b, Dzuba2015c, Yu2016, Yu2018, Yu2019, Yu2022, Beloy2021, Allehabi-PRA-2022, Dzuba2024a, Porsev2024, Yu-PRA-2024, Allehabi-JQSRT-2024} have outlined significant progress in this field. For example, the Ar$^{13+}$ HCI clock has achieved a remarkable systematic frequency uncertainty of $1.6 \times 10^{-18}$ \cite{King2022}, demonstrating the potential of HCIs in high-precision metrology. However, many HCI candidates, particularly those based on heavy elements with complex electronic structures, face substantial experimental challenges, including difficulties in ion preparation, manipulation, and performing high-precision many-body calculations to predict the spectroscopic properties. Additionally, some clock transitions in these ions are weak and challenging to identify experimentally, emphasizing the importance of reliable theoretical predictions for advancing the development of these systems.

The Ir$^{17+}$ ion has attracted significant attention due to its exceptional sensitivity to the fine-structure constant ($\alpha$), which arises from its unique hole configuration in the $4f$ shell and the energy level crossings between the $4f$ and $5s$ orbitals \cite{Berengut2011, Berengut2010}. This high sensitivity positions Ir$^{17+}$ as an ideal candidate for probing variations in fundamental constants and testing other foundational theories. Experimental investigations of M1 transitions in Ir$^{17+}$, conducted using electron beam ion traps (EBITs), have validated the predicted optical transitions stemming from the $5s - 4f$ level crossing \cite{Windberger2015}. Despite extensive theoretical analyses of Ir$^{17+}$’s electronic structure employing many-body theories \cite{Safronova2015, Cheung2020, Dzuba2023}, uncertainties persist in accurately determining its energy levels. Moreover, experimental data on critical properties such as lifetimes, polarizabilities, and hyperfine structure constants remain scarce.

In this paper, the excited energies (EEs) and transition properties of Ir$^{17+}$ ions are investigated using the Kramers-restricted configuration-interaction (KRCI) \cite{Olsen1990, Fleig2003, Fleig2006, Olsen-JCP-2000, Thyssen-JCP-2008, Knecht-JCP-2010} and Fock-space coupled-cluster (FSCC) \cite{Kaldor1991, Visscher-JCP-1996, Visscher-JCP-2002, Pernpointner-JCP-2003, Oleynichenko2020, Oleynichenko2020a} methods within the framework of the Dirac-Coulomb-Gaunt (DCG) Hamiltonian. Our results show multiple forbidden optical transitions between the ground state ($4f^{13}5s$) and the excited states ($4f^{14}$ and $4f^{12}5s^2$). Notably, the $4f^{13}5s~^3F_4^o \rightarrow 4f^{14}~^1S_0$ and $4f^{13}5s~^3F_4^o \rightarrow 4f^{12}5s^2~^3H_6$ transitions, with predicted wavelengths of approximately 346(30) nm and 833(100) nm (from KRCI calculations), exhibit excellent clock features. In particular, the negative differential electric dipole polarizability $\Delta \alpha_d^S$ of the $4f^{13}5s~^3F^o_4 \leftrightarrow 4f^{14}~^1S_0$ transition supports the cancellation of trap-induced Stark and micromotion time-dilation shifts when operated at a “magic” radiofrequency trap drive frequency. Furthermore, these two transitions have opposite relativistic sensitivity coefficients ($K_{\alpha}$), suggesting that their frequency ratio could provide enhanced sensitivity to variations in the fine-structure constant ($\alpha$). Additionally, two other states—$4f^{13}5s~^3F^o_4$ and $4f^{12}5s^2~^3F^o_4$—exhibit lifetimes exceeding 600 ms and 100 ms, respectively, making them promising candidates for clock states as well. When corrections from triple coupled-cluster excitation and finite basis-set size are applied, the FSCC predicted EEs show a clear trend of convergence with the KRCI results. In addition, we present other key properties of the ground and low-lying excited states of Ir$^{17+}$, including Lande $g_J$ factors, lifetimes, electric dipole polarizabilities, electric quadrupole moments, hyperfine structure constants, and isotope shifts. The excellent agreement between the KRCI and FSCC predictions highlights the robustness of our calculations and the reliability of the proposed clock transitions.

\section{Method and Technical Details}

\subsection{The DCG Hamiltonian}
All calculations in this work are performed within the framework of the DCG Hamiltonian. The Hamiltonian, expressed in atomic units, is given by:
\begin{eqnarray}
\hat{H} &=& \sum_i \left[ c (\vec{\bm{\alpha}} \cdot \vec{\bf{p}})_i + (\bm{\beta} - 1)_i m_0 c^2 + V_{\text{nuc}}(r_i) \right] \nonumber \\
&& + \sum_{i<j} \left[ \frac{1}{r_{ij}} - \frac{1}{2} \frac{\vec{\bm{\alpha}}_i \cdot \vec{\bm{\alpha}}_j}{r_{ij}} \right], \label{DCG}
\end{eqnarray}
where $\vec{\bm{\alpha}}$ and $\bm{\beta}$ are the Dirac matrices, $\vec{\bf{p}}$ denotes the momentum operator, $m_0 c^2$ is the electron rest mass energy, and $c$ is the speed of light. The term $V_{\text{nuc}}(r_i)$ represents the nuclear potential, while $r_{ij}$ is the distance between the $i$-th and $j$-th electrons. The final term includes the Gaunt interaction, which is the dominant component of the Breit interaction.

All calculations are performed using the dyall.aaeXz basis sets to investigate the dependence of the results on the basis set size. These basis sets include (25$s$20$p$13$d$9$f$1$g$), (31$s$25$p$16$d$11$f$5$g$1$h$), and (35$s$31$p$20$d$14$f$10$g$5$h$1$i$) functions for $X$ = 2, 3, and 4, respectively \cite{Dyall2023}. These basis sets are specifically designed to optimize the polarization of $d$ shells, account for core and valence electron correlations, and are augmented with diffuse and tight functions tailored for 5$d$ elements. The DHF calculations are implemented in the DIRAC package, which adopts the Gaussian approximation to model the nuclear charge distribution \cite{Visscher-ADNDT-1997}.

When the computational demand is very high, we also employ the Exact Two-Component (X2C) method \cite{Sikkema-JCP-2009, Knecht-JCP-2022}. This approach reduces the four-component Dirac equation to a two-component form, preserving key relativistic effects while significantly enhancing computational efficiency. By combining the X2C method with the DCG Hamiltonian, we perform accurate calculations of key physical properties, such as hyperfine structure constants, polarizabilities, and transition matrix elements. This multi-method approach establishes a robust theoretical framework for investigating the spectroscopic properties of Ir$^{17+}$. 

\subsection{The KRCI calculation}
The Ir$^{17+}$ ion, a heavy system composed of 60 electrons, features a ground state characterized by the $(4f^{13}5s)$ configuration, while low-lying excited states originate from the $(4f^{14})$ and $(4f^{12}5s^2)$ configurations. The near-degeneracy between the open $4f$ and $5s$ orbitals poses significant computational challenges, particularly in accurately treating electron correlation effects. To address these complexities, we employ the KRCI method \cite{Knecht-JCP-2010}, a sophisticated many-body approach well-suited for handling complex multi-electron systems such as Ir$^{17+}$. This method is based on the generalized active space (GAS) formalism \cite{Olsen-JCP-2000, Fleig2001, Fleig2003, Fleig2006}, which allows the inclusion of an arbitrary number of active orbitals with flexible electron occupation constraints, thereby providing a versatile and robust framework for tackling intricate electron correlation problems in open-shell systems.

The KRCI calculation is performed based on an average-of-configuration Hartree–Fock approach (AOC-SCF) \cite{Thyssen-AOC}. The first step involves probing the subtle influence of various DHF reference states on the KRCI energies. We examine several reference configurations: 14 electrons distributed across the $4f$ and $5s$ orbitals (Case I), all 14 electrons placed in the $4f$ orbitals (Case II), 13 electrons in the $4f$ orbitals with 1 electron in the $5s$ orbital (Case III), and 12 electrons in the $4f$ orbitals with 2 electrons in the $5s$ orbital (Case IV). We also consider the multiconfigurational SCF (MCSCF) optimization of the SCF wavefunction \cite{Knecht-JCP-2010}. Upon comparing the energy results for each case, we find that Case II, in which all 14 electrons are assigned to the $4f$ orbitals, yields the most reasonably balanced output for the energy levels of the $4f^{13}5s$, $4f^{14}$, and $4f^{12}5s^{2}$ terms, whereas the other three cases lead to the prediction of incorrect ground states, disordered energy levels, and missed states. The computational details and data analysis, involving extensive tables, are provided in the supplementary materials (SM) \cite{SM}. Therefore, all KRCI calculations in the following are conducted based on the AOC-SCF of Case II.

Although the KRCI method is highly effective for open-shell systems, it encounters significant computational challenges due to the exponential growth in the number of coupled determinants as more active spin-orbitals are included and as the CI excitation ranks expand. To manage this computational complexity while ensuring the accuracy of the results, we truncate the CI model. Unless otherwise stated, we adopt the frozen-core (FC) approximation, wherein the $1s$-$3d$ electrons are treated as inert, while the other orbitals are designated as active. The $4f$ and $5s$ orbitals are considered valence electrons, and core-valence correlation is included by incorporating the $4s$, $4p$, and $4d$ electrons in the correlation treatment using the standard GAS approach.

For computational efficiency, we employ three distinct active spaces in this study. The first model, referred to as e32-CISD, consists of 32 electrons and allows single and double (SD) electron excitation from the $4f$ and $5s$ valence orbitals, while the $4s$, $4p$, and $4d$ core orbitals permit single-electron excitation. This model is implemented using the dyall.aae2z, dyall.aae3z, and dyall.aae4z basis sets to obtain basis-set convergent results. The second model, referred to as e24-CISD, includes SD valence electron excitation from the $4f$ and $5s$ orbitals and single excitation from the $5d$ orbitals. We explore the impact of truncating the virtual orbitals on the computational results using this model with the dyall.aae2z basis set. Furthermore, with all core electrons frozen, we investigate single and double (SD) as well as single, double, and triple (SDT) excitation of the valence electrons, denoted as e14-CISD and e14-CISDT, respectively. The contribution of triple excitation of the valence electrons is estimated based on the energy differences between e14-CISD and e14-CISDT computed using the dyall.aae2z basis set. We present the KRCI results based on the e32-CISD model computed with the dyall.aae4z basis set and estimate the uncertainty by considering possible corrections due to the basis-set size, truncation of the virtual orbital spaces, and triple electronic excitation. The KRCI results under various CI models and basis sets, as well as the uncertainty analysis, are tabulated in the supplementary materials (SM) \cite{SM}.

The atomic properties are calculated by evaluating expectation values over the corresponding single-particle operators using the converged CI wave function. We compute the $g_J$ factor, the E1/M1/E2 transition matrix elements used for calculating the lifetime $\tau$, the hyperfine constant $A$, and the Lorentz-invariance coefficient tensor $T^{(2)}$. All KRCI calculations are performed using the DIRAC program suite \cite{Dirac,Dirac2020}.

\subsection{The FSCC Calculation}
In the FSCC calculation, the reference state is selected as the closed-shell $4f^{14}5s^2$ configuration of Ir$^{15+}$. The coupled-cluster solution for the lowest $(0,0)$ sector is regarded as the ``parent" state, which is subsequently utilized to generate the $(2,0)$ hole sector of the Ir$^{17+}$ ion by removing two electrons. The $4f5s$ constitute the smallest model space sufficient to yield our target states: the $4f^{13}5s$ ground state, obtained by removing one electron from the $4f$ and another from the $5s$ spinors; the $4f^{14}$ state, formed by subtracting two electrons from the $5s$ spinors; and the $4f^{12}5s^2$ state, generated by removing two electrons from the $4f$ spinors. Expanding the model space to include additional shells, such as $4d$ and $4s4p$, in the active space is found to have a negligible influence on the computational results. Since our target states originate from the hole sector, the particle sector is deemed weakly relevant.

We consider all 60 electrons in the CC calculation at the single- and double-excitation (SD) levels and include virtual orbitals with energy above 100 a.u., referred to as e60-CCSD. These calculations are conducted for the dyall.aae2z, dyall.aae3z, and dyall.aae4z basis sets. We have verified that truncation of the virtual orbitals has a negligible impact on the results. While a full FSCC SDT calculation provides higher accuracy, its computational cost is prohibitively high. The full FSCC SDT calculation is crucial for the Ir$^{17+}$ ion; however, this method is highly computationally demanding. Therefore, we employ this method with the dyall.aae2z basis set with truncation of virtual orbitals with energy less than 20 a.u., excluding the innermost 28 core electrons from the correlation treatment. This calculation is referred to as ``e32-CCSDT-$2\xi$". Due to resource constraints, we cannot perform the full FSCC SDT calculation with larger basis sets; however, we have investigated the influence of excluding the inner core on the computational results at the CCSD level, which is found to be small.

We have implemented the FSCC calculation using two publicly available versions of the FSCC code: one from the DIRAC package \cite{Visscher-JCP-1996, Visscher-JCP-2002, Pernpointner-JCP-2003} and the other from the EXP-T code \cite{Oleynichenko2020, Oleynichenko2020a}. The EXP-T code offers an advantage for implementing triple-rank excitation. The FSCC calculation is performed following the DHF calculation of the Ir$^{15+}$ ion and the subsequent integral transformation of the molecular spinors, both of which are accomplished using the DIRAC package. The exact two-component (X2C) method \cite{Peng2013, Knecht-JCP-2022} is employed in the FSCC calculations. Given the exceptional performance of the EXP-T code in handling triple-rank excitation, we employ its FSCC module to perform high-precision energy calculations for the  Ir$^{17+}$ ion. The results are systematically structured and thoroughly analyzed. Due to computational resource constraints, finite-field calculations for  Ir$^{17+}$ ion properties are conducted using the FSCC module within the DIRAC software package. To ensure reliability and accuracy within the available computational capacity, the EXPT procedure is employed for cross-validation of the FSCC results. A rigorous comparative analysis demonstrates a high level of consistency between the energy calculations and property results obtained from the DIRAC and EXPT programs at the CCSD level. This key finding not only confirms the robustness and stability of the adopted computational methodologies but also provides a solid foundation for further high-precision investigations in this field.
	
\begin{table*}[htp]\caption{Energy levels (in cm$^{-1}$) of the Ir$^{17+}$ ion. The KRCI (e32-CISD) calculation and FSCC (e60-CCSD) calculations consider the influence of basis set size, truncation of virtual orbitals, and triple excitation. Analysis of uncertainties (given in parentheses) is provided in the SM \cite{SM}.\label{tab:Ir17+_energy}}{\setlength{\tabcolsep}{6pt}
		\begin{tabular}{ccc ccc ccc ccc}\hline\hline  \addlinespace[0.1cm]
\multirow{2}{*}{Config.}	&	\multirow{2}{*}{Term}	&	\multirow{2}{*}{$g$-factor}	&&	\multicolumn{5}{c}{CI}									&&	\multicolumn{2}{c}{CC}			\\ \cline{5-9}  \cline{11-12} \addlinespace[0.1cm]
&		&		&&KRCI	&	Ref.\cite{Berengut2011}	&	Ref.\cite{Safronova2015}	&	Ref.\cite{Cheung2020}	&	Ref.\cite{Dzuba2023}	&&	FSCC	&	Ref.\cite{Windberger2015}	\\\hline \addlinespace[0.1cm]
$4f^{13}5s$	&	$^3F_4^o$	&	1.2477 	&&	0	&	0	&	0	&	0	&	0	&&	0 	&	0	\\
&	$^3F_3^o$	&	1.0490 	&&	4769(45)	&	4838	&	4236	&	4777	&	4647	&&	4655 (74)	&	4662 	\\
&	$^3F_2^o$	&	0.6642 	&&	26194(26)	&	26272 	&	26174	&	25186 	&	25198 	&&	25476(129)	&	25156 	\\
&	$^1F_3^o$	&	1.0296 	&&	31381(109)	&	31492 	&	30606	&	30395 	&	30167 	&&	30579(232)	&	30197 	\\ \addlinespace[0.1cm]
$4f^{14}$	&	$^1S_0$	&	0.0000 	&&	12006(1392)	&	5055 	&	5091	&	12382 	&	7424 	&&	10203(3563)	&	13599 	\\ \addlinespace[0.1cm]
$4f^{12}5s^2$	&	$^3H_6$	&	1.1612 	&&	28848(2117)	&	35285 	&	33856	&	30283 	&	29695 	&&	27445(3007)	&	24221 	\\
&	$^3F_4$	&	1.1359 	&&	38221(2019)	&	45214 	&	42199	&	39564 	&	39563 	&&	36062(2774) 	&	33545 	\\
&	$^3H_5$	&	1.0309 	&&	53162(2115)	&	59727 	&	58261	&	53798 	&	53668 	&&	51059(3120) 	&	47683 	\\
&	$^3F_2$	&	0.8426 	&&	60530(1509)	&	68538 	&	63696	&	61429 	&	62140 	&&	56931(2523) 	&	55007 	\\
&	$^1G_4$	&	0.9927 	&&	61953(2058)	&	68885 	&		&	62261 	&	62380 	&&	59318(2977) 	&	56217 	\\
&	$^3F_3$	&	1.0809 	&&	64577(1795)	&	71917 	&	68886	&	65180 	&	65438 	&&	61492(2886) 	&	58806 	\\
&	$^3H_4$	&	0.9141 	&&	85218(2150)	&	92224 	&	66296	&	84524 	&	84662 	&&	82019(3066) 	&	78534 	\\
&	$^1D_2$	&	1.1313 	&&	89257(1064)	&	98067 	&	117322	&	89273 	&	91341 	&&	84415(2351) 	&	82422 	\\
&	$^1J_6$	&	1.0005 	&&	100121(688)	&	110065 	&		&	101136 	&	103487 	&&	96331(2120) 	&	93867 	\\ \hline \hline  
	\end{tabular}}
\end{table*}

\subsection{The finite-field Calculation}
The field-field (FF) approach provides a convenient method for calculating electric-field response properties, such as the static electric-dipole scalar and tensor polarizabilities $\alpha_d^{S,T}$ and the electric quadrupole moment $\Theta$ \cite{Archibong1991, Guo-PRA-2021}. This approach is also employed for calculating hyperfine structure constants \cite{Haase-JPCA-2020, Denis-PRA-2022} and the isotope shift. 
In quantum chemical calculations, the FF method offers distinct advantages. With a concise algorithm grounded in quantum mechanical perturbation theory, the FF method effectively accounts for the wave function’s response to the external field and its influence on molecular properties. This advantage holds true regardless of whether the CC method is variational. In contrast, the expectation value method only considers the static terms of the system, neglecting dynamic effects caused by the wave function’s response to the external field.

In the calculation of the static electric dipole polarizability $\alpha_d^{S,T}$, a finite electric field perturbation is introduced into the Hamiltonian via the term $\vec{D} \cdot \mathcal{\vec{E}}$, where $\vec{D}$ denotes the electric dipole operator and $\mathcal{\vec{E}}$ represents the applied electric field. By varying the field strength $|\mathcal{\vec{E}}|$, energy values are computed and fitted to a Taylor expansion centered at $|\mathcal{\vec{E}}| = 0$. The second term of this expansion yields the electric dipole polarizability $\alpha_d(J_n, M_{J_n})$ for all possible $M_{J_n}$ values, from which the static polarizabilities $\alpha_d^S$ and $\alpha_d^T$ are derived. The computation of the electric quadrupole moment $\Theta$ follows a similar approach. An interaction Hamiltonian of the form $\vec{\Theta} \cdot \vec{\nabla} |\mathcal{\vec{E}}|$ is introduced, where $\vec{\nabla} |\mathcal{\vec{E}}|$ represents the gradient of the electric field. As the gradient $\vec{\nabla} |\mathcal{\vec{E}}|$ is varied, the system’s energy exhibits a linear dependence. By fitting this linear relationship, the coefficient of the fit provides the value of $\Theta$, which is defined as the second-rank tensor describing the distribution of charge within the system. This tensor encodes information about the asymmetry of the charge distribution and plays a critical role in understanding the interaction of the system with external electric field gradients.

The FF calculation of the hyperfine structure constant $A$ is implemented in the DIRAC program, where the nuclear magnetic-dipole operator is incorporated into the FSCC calculation step, as described in Refs. \cite{Haase-JPCA-2020, Denis-PRA-2022}. For the FF calculation of the field shift, we adopt one-electron operators for Fermi-contact integrals corresponding to a specific nucleus, which are added to the DHF Hamiltonian. The electron density at the nucleus is obtained by fitting the finite-field energies. The field-shift factor $F$ is evaluated as $F = Ze^2 \Delta \rho_0 / 6\epsilon_0$, where $Z$ is the nuclear charge number, $e$ is the elementary charge, $\epsilon_0$ is the vacuum permittivity, and $\Delta \rho_0$ represents the differential nuclear charge density for a transition.

\begin{table*}[htp]
	\caption{The lifetime ($\tau$), the relativistic enhancement factor ($K_\alpha$), and the Lorentz invariance violations (LLI) reduced matrix elements ($ \langle J \|T^{(2)}\|J\rangle$) are obtained using the e32-CISD calculation with the dyall.aae4z basis set, along with comparisons to previous theoretical studies. The error bar in the $\tau$ value (given in parentheses) accounts for the uncertainty in the energies calculated by the KRCI method. \label{tab:Ir17+_property}}{\setlength{\tabcolsep}{10pt}
		\begin{tabular}{cccccccccccc}\hline\hline  \addlinespace[0.1cm]
			\multirow{2}{*}{Config.}	&	\multirow{2}{*}{Term}	&	\multicolumn{2}{c}{$\tau$ (ms)}			&&	\multicolumn{2}{c}{$K_\alpha$}			&&	\multicolumn{2}{c}{$ \langle J \|T^{(2)}\|J\rangle  $}			\\ \cline{3-4} \cline{6-7} \cline{9-10}\addlinespace[0.1cm]
			&		&	e32-CISD	&	Ref.\cite{Cheung2020}	&&	e32-CISD	&	Ref.\cite{Berengut2011}	&&	e32-CISD	&	Ref.\cite{Dzuba2023}	\\\hline \addlinespace[0.1cm]
			$4f^{13}5s$	&	$^3F_4^o$	&		&		&&		&		&&	-263 	&	-283	\\
			&	$^3F_3^o$	&	610(48)	&	849 	&&	0.7 	&	0.9 	&&	-217 	&	-233	\\
			&	$^3F_2^o$	&	4.60(4)	&	4.42 	&&	1.9 	&	1.9 	&&	-183 	&	-197	\\
			&	$^1F_3^o$	&	3.24(8)	&	3.27 	&&	1.6 	&	1.7 	&&	-236 	&	-254	\\ \addlinespace[0.1cm]
			$4f^{14}$	&	$^1S_0$	&	--	&		&&	68.7 	&	145.3 	&&	0 	&	0	\\ \addlinespace[0.1cm]
			$4f^{12}5s^2$	&	$^3H_6$	&	--	&		&&	-35.5 	&	-26.0 	&&	-291 	&	-311	\\
			&	$^3F_4$	&	104(7)	&	14.90 	&&	-27.6 	&	-20.0 	&&	26 	&	26	\\
			&	$^3H_5$	&	2.662(1)	&	2.63 	&&	-17.7 	&	-13.0 	&&	-249 	&	-266	\\
			&	$^3F_2$	&	48.8(4)	&	2.79 	&&	-16.7 	&	-12.0 	&&	127 	&	131	\\
			&	$^1G_4$	&	1.82(5)	&	2.47 	&&	-15.6 	&	-12.0 	&&	-95 	&	-107	\\
			&	$^3F_3$	&	1.58(4)	&	3.13 	&&	-14.8 	&	-11.0 	&&	67 	&	71	\\
			&	$^3H_4$	&	0.54(2)	&	0.66 	&&	-10.9 	&	-8.0 	&&	-134 	&	-138	\\
			&	$^1D_2$	&	1.53(14)	&	0.28 	&&	-10.6 	&	-8.0 	&&	87 	&	95	\\
			&	$^1J_6$	&	4.58(54)	&		&&	-9.1 	&	-7.0 	&&	-576 	&	-612	\\ \hline \hline  
	\end{tabular}}
\end{table*}

\section{Results and Discussion}

\subsection{Energy level}
Table \ref{tab:Ir17+_energy} presents the energy levels and corresponding $g$-factors of the Ir$^{17+}$ ion obtained through KRCI and FSCC calculations, along with a comprehensive comparison to other CI- and CC-type theoretical data. The $4f^{14}~^1S_0$ state draws significant attention due to its substantial positive $\alpha$-sensitivity, which distinguishes it from other states in the Ir$^{17+}$ ion. Prior predictions for the energy of this state show considerable variance, ranging from 5000 to 7000 cm$^{-1}$ \cite{Berengut2011, Safronova2015, Dzuba2023} to 12000 cm$^{-1}$ \cite{Cheung2020}. Our KRCI result, 12006 cm$^{-1}$, demonstrates an agreement with the latter prediction \cite{Cheung2020}. Similarly, for another promising clock state, $4f^{12}5s^2~^3H_6$, the KRCI results align closely with the predictions from Ref.\cite{Cheung2020} and Ref.\cite{Dzuba2023}.

For the ground state $4f^{13}5s^1$, the uncertainties in the KRCI results are minimal, within 100 cm$^{-1}$, while for the low-lying excited states $4f^{14}$ and $4f^{12}5s^2$, the uncertainties are as large as 2000 cm$^{-1}$. These uncertainties primarily stem from computational instability when applying the truncated CI model with different Dyall basis sets, especially affecting high-energy configurations. A comparison of e32-CISD energy values calculated with the dyall.aae3z and dyall.aae4z basis sets reveals that the dyall.aae3z set tends to underestimate energy values, leading to large uncertainties due to the basis set size. 
The analysis of the uncertainty in KRCI calculation is presented in Table IV of SM \cite{SM}. Our analysis indicates that the primary source of uncertainty currently arises from the incompleteness of the basis set. As Gaussian basis sets continue to improve, we anticipate a substantial reduction in this uncertainty.
The excellent concordance between the KRCI results and several recent CI calculations \cite{Cheung2020, Dzuba2023} validates the method’s robustness. At its current level (e32-CISD with the dyall.aae4z basis set), the KRCI method achieves a remarkable balance between computational efficiency and accuracy, providing an optimal framework for tackling such complex systems.

In the investigation of the Ir$^{17+}$ ion, the FSCC method demonstrates remarkable performance, particularly in the calculation of low-energy states. For instance, the predicted values for the $4f^{13}5s~^3F_3^o$, $^3F^o_2$, and $^1F_3$ states align exceptionally well with those obtained from CI methods. Nonetheless, discrepancies persist between the FSCC results and CI predictions. The earlier FSCC prediction for the $4f^{14}~^1S_0$ state, as reported in Ref. \cite{Windberger2015}, is significantly higher than both the CI results \cite{Cheung2020, Dzuba2023}. For the excited states of the $4f^{12}5s^2$ configuration, FSCC-calculated energy values are generally lower than the corresponding CI results. In the current work, we conduct the FSCC calculation using a series of progressively larger Gaussian basis sets. The dyall.aae4z basis set utilized in this study is comparable to the even-tempered basis set from Ref. \cite{Windberger2015}. Our e60-CCSD calculations reveal a trend of increasing energy values with larger basis sets. Specifically, when transitioning from the dyall.aae3z to the dyall.aae4z basis set, we observe an energy increase of the $4f^{12}5s^2$ states by over 1000 cm$^{-1}$, indicating that full convergence has not yet been achieved. We anticipate that FSCC energy values will continue to rise with the adoption of higher-order basis sets. Furthermore, we implement FSCC SDT calculations. With the inclusion of the triple contribution, the FSCC prediction for the energy of the $4f^{14}~^1S_0$ state decreases significantly, while those for the $4f^{12}5s^2$ states increase by more than 2000 cm$^{-1}$. This indicates that if we consider possible corrections due to basis-size expansion and triple excitation, the FSCC results are expected to align with the KRCI results. By incorporating these corrections arising from both the larger basis set and the inclusion of SDT excitation. our FSCC results show significantly improved agreement with the CI predictions from Refs. \cite{Cheung2020} and \cite{Dzuba2023}, as well as with our own CI results. We acknowledge that the uncertainty of our current FSCC-calculated values remains rather large, which is based on conservative estimates of the uncertainties in the triple-excitation and basis-size expansion corrections, as explained in the SM \cite{SM}. Notably, the current FSCC SDT calculation is still performed with the dyall.aae2z basis set due to practical computational constraints. We expect to suppress such uncertainty by performing FSCC SDT calculations with larger basis sets in the future.

\subsection{$\tau$, $K_{\alpha}$, $\langle J \| T^{(2)} \| J \rangle$}

Table \ref{tab:Ir17+_property} presents the theoretical results for the lifetimes ($\tau$), sensitivity coefficients ($K_{\alpha}$), and Lorentz invariance violations (LLI) reduced matrix elements ($\langle J \| T^{(2)} \| J \rangle$) for the Ir$^{17+}$ ion, calculated using the KRCI method with the e32-CISD model. The lifetime calculation utilizes the Einstein spontaneous emission rate. The detailed formulas and specific numerical values are provided in the SM \cite{SM}. For the low-energy states within the $4f^{13}5s$ configuration, such as the $^3F^o_4$ state, we determine a lifetime of $610(48)$ ms, which exhibits a moderate discrepancy compared to the value of $849$ ms reported in Ref. \cite{Safronova2015}. However, the lifetimes of states such as $^3F^o_3$ and $^1F^o_3$, with values of $4.60(4)$ ms and $3.24(8)$ ms, respectively, show much better agreement with the literature, highlighting the precision of our electron correlation treatment for these low-energy states.

\begin{table*}[hbt]
	\caption{The scalar ($\alpha_d^S$) and tensor ($\alpha_d^T$) electric dipole polarizabilities, as well as the differential polarizability ($\Delta \alpha_d^S$) between the excited and ground states, are presented. The analysis of uncertainties (given in parentheses) is provided in the SM \cite{SM}.\label{tab:Ir17+_alpha}}{\setlength{\tabcolsep}{6pt}
		\begin{tabular}{ccc ccc ccc c}\hline\hline  \addlinespace[0.1cm]
			\multirow{2}{*}{Config.}	&	\multirow{2}{*}{Term}	&&	\multicolumn{3}{c}{e32-CISD}					&&	\multicolumn{3}{c}{e60-CCSD}					\\ \cline{4-6}  \cline{8-10} \addlinespace[0.1cm]
			&		&&	$\alpha_d^S$	&	$\alpha_d^T$	&	$\Delta \alpha_d^S$	&&	$\alpha_d^S$	&	$\alpha_d^T$	&	$\Delta \alpha_d^S$	\\\hline \addlinespace[0.1cm]
			$4f^{13}5s$	&	$^3F_4^o$	&&	0.3600(227)	&	0.0085(12)	&	0.0000(0)	&&	0.3736(24)	&	0.0077(64)	&	0.0000(0)	\\
			&	$^3F_3^o$	&&	0.3607(217)	&	0.0046(20)	&	0.0007(9)	&&	0.3744(1)	&	0.0133(209)	&	0.0008(24)	\\
			&	$^3F_2^o$	&&	0.3591(220)	&	0.0043(21)	&	-0.0009(7)	&&	0.3731(0)	&	0.0161(74)	&	-0.0005(24)	\\
			&	$^1F_3^o$	&&	0.3604(226)	&	0.0025(12)	&	0.0004(1)	&&	0.3747(59)	&	0.0080(149)	&	0.0011(35)	\\ \addlinespace[0.1cm]
			$4f^{14}$	&	$^1S_0$	&&	0.1570(466)	&	0	&	-0.2030(239)	&&	0.1734(119)	&	0	&	-0.2002(96)	\\ \addlinespace[0.1cm]
			$4f^{12}5s^2$	&	$^3H_6$	&&	0.4850(127)	&	0.0106(4)	&	0.1250(10)	&&	0.4979(27)	&	0.0079(30)	&	0.1243(4)	\\
			&	$^3F_4$	&&	0.4853(64)	&	-0.0011(66)	&	0.1253(162)	&&	0.4980(10)	&	0.0105(538)	&	0.1244(33)	\\
			&	$^3H_5$	&&	0.4842(126)	&	0.0101(3)	&	0.1242(101)	&&	0.4984(30)	&	0.0072(34)	&	0.1248(6)	\\
			&	$^3F_2$	&&	0.4846(126)	&	-0.0059(19)	&	0.1246(100)	&&	0.4976(27)	&	-0.0140(276)	&	0.1240(3)	\\
			&	$^1G_4$	&&	0.4845(137)	&	0.0042(3)	&	0.1245(90)	&&	0.4984(11)	&	0.0108(135)	&	0.1248(13)	\\
			&	$^3F_3$	&&	0.4845(135)	&	-0.0021(14)	&	0.1245(92)	&&	0.4989(11)	&	-0.0032(290)	&	0.1253(35)	\\
			&	$^3H_4$	&&	0.4835(131)	&	0.0048(18)	&	0.1235(96)	&&	0.4967(37)	&	0.0140(211)	&	0.1231(14)	\\
			&	$^1D_2$	&&	0.4839(125)	&	-0.0033(4)	&	0.1239(102)	&&	0.4985(2)	&	-0.0086(169)	&	0.1249(26)	\\
			&	$^1J_6$	&&	0.4842(136)	&	0.0203(20)	&	0.1242(90)	&&	0.4982(10)	&	0.0161(1)	&	0.1245(34)	\\ \hline \hline  
	\end{tabular}}
\end{table*}

The strongly forbidden transition between the $4f^{14}~^1S_0$ state and the $4f^{12}5s^2~^3H_6$ state, along with the ground state, indicates that these states effectively possess very long lifetimes. For the $4f^{12}5s^2~^3F_4$ state, we calculate a lifetime of $104(7)$ ms, which differs significantly from the earlier prediction of $14.90$ ms reported in Ref. \cite{Safronova2015}. This substantial discrepancy arises because our KRCI calculations predict the E1 transition between the $4f^{12}5s^2~^3F_4$ state and the ground states ($4f^{13}5s~^3F_4^o$ and $4f^{13}5s~^{(3,1)}F_3^o$) to be considerably weaker, by an order of magnitude, than those reported in Ref. \cite{Safronova2015}. For the transition probabilities of $(4f^{12}5s^2)~^3F_4 \rightarrow (4f^{13}5s)~^3F_3^o$ and $(4f^{13}5s)~^1F_3^o$, our calculations yield values of $45.6$ and $4.00 \times 10^{-4}$, respectively, whereas Ref. \cite{Safronova2015} reports values of $6.51 \times 10^1$ and $1.79$. More detailed data are provided in the SM \cite{SM}. Despite the discrepancies in certain data, our results for other states show remarkable agreement with the literature, further validating the overall reliability and consistency of our findings.

\begin{table*}[hbt]
	\caption{The electric quadrupole moment ($\Theta$), hyperfine structure constant ($A$), and field shift factor ($F$) for the transitions between the excited and ground states are presented. The analysis of uncertainties (given in parentheses) is provided in the SM \cite{SM}.\label{tab:Ir17+_TAF}}{\setlength{\tabcolsep}{8pt}
		\begin{tabular}{ccc ccc ccc cc}\hline\hline  \addlinespace[0.1cm]
\multirow{2}{*}{Config.}	&	\multirow{2}{*}{Term}	&&	\multicolumn{2}{c}{$\Theta$ a.u.}			&&	\multicolumn{2}{c}{$A$ MHz}			&&	$F$ (GHz/fm$^2$)	&	\\ \cline{4-5}  \cline{7-8}  \cline{10-11} \addlinespace[0.1cm]
	&		&&	e32-CISD	&	e60-CCSD	&&	e32-CISD	&	e60-CCSD	&&	e60-CCSD	&	\\\hline \addlinespace[0.1cm]
$4f^{13}5s$	&	$^3F_4^o$	&&	-0.0870(36)	&	-0.0853(15)	&&	5193(60)	&	5268(2)	&&	0	&	\\
	&	$^3F_3^o$	&&	-0.0750(30)	&	-0.0738(12)	&&	-4325(61)	&	-4434(13)	&&	-1.920(10)	&	\\
	&	$^3F_2^o$	&&	-0.0586(24)	&	-0.0576(10)	&&	-6210(81)	&	-6346(4)	&&	1.44(10)	&	\\
	&	$^1F_3^o$	&&	-0.0796(34)	&	-0.0788(14)	&&	6564(82)	&	6676(11)	&&	-1.97(20)	&	\\ \addlinespace[0.1cm]
$4f^{14}$	&	$^1S_0$	&&	0	&	0	&&	0	&	0	&&	-617(2)	&	\\ \addlinespace[0.1cm]
$4f^{12}5s^2$	&	$^3H_6$	&&	-0.0822(23)	&	-0.0798(11)	&&	245(2)	&	219(2)	&&	643(5)	&	\\
	&	$^3F_4$	&&	0.0091(10)	&	0.0095(2)	&&	242(2)	&	219(1)	&&	643(4)	&	\\
	&	$^3H_5$	&&	-0.0722(21)	&	-0.0702(10)	&&	296(1)	&	279(1)	&&	644(4)	&	\\
	&	$^3F_2$	&&	0.0398(16)	&	0.0392(6)	&&	241(2)	&	254(1)	&&	645(6)	&	\\
	&	$^1G_4$	&&	-0.0294(18)5	&	-0.0291(1)	&&	293(1)	&	281(1)	&&	645(4)	&	\\
	&	$^3F_3$	&&	0.0201(12)	&	0.0197(4)	&&	271(1)	&	252(1)	&&	645(4)	&	\\
	&	$^3H_4$	&&	-0.0383(10)	&	-0.0372(4)	&&	379(1)	&	366(3)	&&	646(4)	&	\\
	&	$^1D_2$	&&	0.0252(16)	&	0.0246(6)	&&	270(2)	&	244(2)	&&	645(4)	&	\\
	&	$^1J_6$	&&	-0.1565(50)	&	-0.1523(25)	&&	298(1)	&	286(1)	&&	645(5)	&	\\ \hline \hline  
	\end{tabular}}
\end{table*}

The enhancement factor ($K_\alpha$), which quantifies the strength of spin-orbit coupling and the quality of electron correlation treatments, provides critical insights into the system’s behavior. Ir$^{17+}$ ions exhibit exceptional sensitivity to variations in the fine-structure constant $\alpha$ ($\alpha = \frac{e^2}{\hbar c}$), a characteristic that supports their proposed role in precision measurements, as detailed in \cite{Berengut2011, Berengut2010}. In the context of atomic transitions, the sensitivity to changes in $\alpha$ is maximized for transitions involving significant changes in total electronic angular momentum $J$ within the single-electron approximation, as discussed in \cite{Berengut2011}. The Ir$^{17+}$ ion undergoes optical transitions between the $4f^{12}5s^2$ and $4f^{13}5s$ configurations, corresponding to single-electron $s \leftrightarrow f$ transitions with $\Delta J$ values of 2 or 3. For low-energy states in the $4f^{13}5s$ configuration, such as the $^3F^o_4$ state ($K_\alpha = 0.7$) and the $^3F^o_3$ state ($K_\alpha = 1.9$), the computed values align well with the results presented in \cite{Berengut2011}, thereby validating the theoretical predictions. However, for higher-energy states such as $^1S_0$ and $^3H_6$, the $K_\alpha$ values are computed to be $68.7$ and $-35.5$, respectively. The observed trend of a shift in $K_\alpha$ values from positive to negative with increasing excitation energy demonstrates an enhanced sensitivity to $\alpha$ variation when using the frequency ratio of the $^3F_4 \leftrightarrow ^1S_0$ and $^3F_4 \leftrightarrow ^3H_6$ transitions.

The normalized matrix elements ($\langle J \| T^{(2)} \| J \rangle$) play a critical role in the investigation of Lorentz invariance violations (LLI). As noted in \cite{Shaniv2018}, a promising approach for probing LLI violations involves measuring the transition frequencies between states with distinct projections of the total electronic angular momentum $J_z$ within a single metastable state. High sensitivity is achieved when the $T^{(2)}$ operator matrix element is large and the state in question exhibits a long lifetime. The KRCI results presented herein demonstrate excellent agreement with the findings reported in \cite{Dzuba2023}. For Ir$^{17+}$, the ground state and excited states (excluding the $4f^{14}~^1S_0$ state) all exhibit substantial matrix elements for the ($\langle J \| T^{(2)} \| J \rangle$) operator. These values are significantly larger than those observed in Yb$^+$ ions, which are widely regarded as leading candidates for the most sensitive LLI measurements. Collectively, these findings underscore the promising potential of Ir$^{17+}$ ions as an exceptional system for investigating LLI violations.

\subsection{$\alpha_d^{S, T}$}
Table \ref{tab:Ir17+_alpha} provides a comprehensive analysis of the electric dipole polarizabilities for both the ground and low-lying excited states of the Ir$^{17+}$ ion, calculated using the FF method within the KRCI and FSCC frameworks. Despite differences in computational strategies, the results for both the scalar electric dipole polarizability ($\alpha_d^S$) and the tensor electric dipole polarizability ($\alpha_d^T$) demonstrate remarkable agreement, thereby offering robust evidence for the numerical consistency and reliability of both approaches.

For each excited state, the differential values $\Delta \alpha_d^S$ are calculated with respect to the ground state, $4f^{13}5s~^3F^o_4$. Of particular note is the negative $\Delta \alpha_d^S$ between the $4f^{14}~^1S_0$ and $4f^{13}5s~^3F^o_4$ states. This negative value is useful for realization of a “magic” RF trap frequency $\nu_{\text{magic}}$, a frequency that effectively negates both the DC Stark shift and the micromotion shift, as discussed in \cite{Dube-PRA-2013}. Using the simplified formula $\nu_{\text{magic}} = -\frac{h\nu}{\Delta \alpha_d^S}(\frac{q}{2\pi m c})^2$, where $q = 17$ denotes the charge of the Ir$^{17+}$ ion, $m$ is the mass of the Ir atom, $\nu$ is the transition frequency, and $h$ and $c$ are Planck’s constant and the speed of light, respectively, the calculated $\nu_{\text{magic}}$ is estimated to be approximately $500 \times 2\pi$ MHz, which falls within a reasonable range for typical experimental trap configurations.

Furthermore, the $4f^{14}~^1S_0$ state exhibits zero tensor dipole polarizability. For the $4f^{12}5s^2$ configurations, the $\Delta \alpha_d^S$ value for the $^3H_6$ state is positive, indicating that it does not support the cancellation of the Stark shifts and micromotion time-dilation shifts induced by the trap. However, the relatively small values of $\Delta \alpha_d^S$ and $\alpha_d^T$ (approximately 0.04) suggest a substantial suppression of both Stark shifts and black-body radiation (BBR) shifts, thereby ensuring minimized systematic errors in precision measurements.

\subsection{$\Theta$, $A$ and $F$}

Table \ref{tab:Ir17+_TAF} provides a detailed analysis of the electric quadrupole moment ($\Theta$) for both the ground and low-lying excited states of the Ir$^{17+}$ ion. The $\Theta$ values, obtained as expectation values of the single-electron operator in the e32-CISD calculation and the e60-CCSD FSCC calculation using the FF approach, demonstrate excellent agreement. Notably, the $4f^{14}~^1S_0$ state is characterized by $\Theta = 0$, owing to its total angular momentum $J = 0$, whereas states in the $4f^{13}5s$ and $4f^{12}5s^2$ configurations exhibit non-zero $\Theta$ values. The significance of these $\Theta$ values lies in their impact on the interaction between the quadrupole moment of an atomic state and an external electric field gradient, which can lead to residual quadrupole shifts in clock transitions. Particularly noteworthy is the fact that both the $4f^{13}5s~^3F_4^o$ ground state and the $4f^{12}5s^2~^3H_6$ excited state possess extremely small $\Theta$ values, making them highly advantageous for optical clock applications due to the substantial suppression of electric quadrupole shifts. Furthermore, the $\Theta$ value for the $4f^{12}5s^2$ configuration is an order of magnitude smaller than that of the other states, further affirming its suitability for clock transitions.

Additionally, Table \ref{tab:Ir17+_TAF} presents the hyperfine structure constants ($A$) for the ground and low-lying excited states of Ir$^{17+}$. The results from both e32-CISD expectation and e60-CCSD FF calculations demonstrate a high degree of consistency, with discrepancies ranging from only 2-10\%, highlighting the robustness of the computational approach. The ground state and states in the $4f^{13}5s$ configuration exhibit relatively large $A$ values, while the $4f^{12}5s^2$ states, though exhibiting smaller constants, display notable stability. For instance, the $^3H_6$ state has hyperfine constants of $245(2)$ and $219(2)$, which indicate weaker but stable nuclear-electronic interactions, attributed to a single-electron excitation from the $4f$ to the $5s$ orbital. In the $4f^{14}$ configuration, the $^1S_0$ state has a zero hyperfine structure constant, consistent with its symmetric electronic distribution, further reinforcing its potential as a stable reference for optical clock applications. The comparison between the KRCI and FSCC methods demonstrates a high level of consistency in predictions, validating their reliability across different energy states.

Moreover, the field-shift factor ($F$) associated with the isotope shift of the Ir$^{17+}$ ion is calculated using the e60-CCSD method in conjunction with the FF approach. The $F$ values for transitions between the $4f^{13}5s~^3F_4^o$ ground state and the $4f^{14}~^1S_0$ state, as well as for transitions between the same ground state and the $4f^{12}5s^2$ configurations, exhibit similar magnitudes but opposite signs. This intriguing observation paves the way for future investigations of Ir$^{17+}$, particularly in the realm of isotope nonlinear field displacement and nuclear structure, where novel insights may be gained.

\section{Conclusions}
This study provides a comprehensive theoretical exploration of the atomic properties of Ir$^{17+}$, offering critical insights into its prospective applications in optical clock technology and fundamental physics research. Within the framework of the DCG Hamiltonian and utilizing both the KRCI and FSCC methods, we calculate the energy and transition properties of the Ir$^{17+}$ ion. The energy level computations demonstrate that the KRCI method is particularly effective for predicting high-lying states, while the FSCC method excels in the treatment of low-lying states. Although some discrepancies arise between these two methods and previous studies, incorporating higher-order excitation and employing larger basis sets significantly enhance the precision of the results. For instance, the KRCI results for the $4f^{14}~^1S_0$ state show strong agreement with certain prior predictions, while FSCC, after accounting for triple excitation and utilizing a more extensive basis set, tends to agree with CI-based results.

Regarding other properties, states such as $4f^{13}5s~{^3F_3^o}$ and $4f^{12}5s^2~{^3F_4}$, which exhibit relatively long lifetimes, emerge as promising candidates for optical clock transitions. The Ir$^{17+}$ ion demonstrates remarkable sensitivity to variations in the fine-structure constant $\alpha$, and certain $K_{\alpha}$ values suggest that further refinement is necessary to fully capture electronic correlation effects, particularly in high-energy states. In terms of polarizabilities, the negative $\Delta\alpha_d^S$ value between the $4f^{14}~^1S_0$ and $4f^{13}5s~{^3F_4^o}$ states enables the realization of a “magic” radio-frequency trap drive frequency, which is critical for mitigating both Stark shifts and micromotion time dilation effects. The remarkably small electric quadrupole moments of the $4f^{13}5s~{^3F_4^o}$ ground state and the $4f^{12}5s^2~{^3H_6}$ excited state contribute significantly to the suppression of electric quadrupole shifts, making these states highly advantageous for optical clock applications. The hyperfine structure constants calculated using different methods exhibit exceptional consistency, while the field-shift factor $F$ opens promising avenues for future research on Ir$^{17+}$, particularly in isotope-dependent studies.

In conclusion, this investigation confirms the reliability of the KRCI and FSCC methods for probing the properties of Ir$^{17+}$. The identified potential optical clock transitions, such as $4f^{13}5s~{^3F_4^o} \to 4f^{14}~^1S_0$ and $4f^{13}5s~{^3F_4^o} \to 4f^{12}5s^2~{^3H_6}$, along with the exceptional characteristics of the corresponding states, highlight the extraordinary potential of Ir$^{17+}$ in the domain of optical clocks. This work establishes a robust theoretical foundation for future experimental endeavors on Ir$^{17+}$ and paves the way for the exploration of fundamental physics, such as testing the variation of fundamental constants.

\section{ACKNOWLEDGMENTS}
This work is supported by Innovation Program for Quantum Science and Technology (2021ZD0300901), the National Key Research and Development Program of China (2021YFA1402104), and the Space Application System of the China Manned Space Program.



\begin{thebibliography}{}
\bibitem{Brewer-PRL-2019}
S. M. Brewer, J. -S. Chen, A. M. Hankin, E. R. Clements, C. W. Chou, D. J. Wineland, D. B. Hume, and D. R. Leibrandt, Phys. Rev. Lett. {\bf 131}, 059901 (2023).

\bibitem{Lange-PRL-2021}
R. Lange, A. A. Peshkov, N. Huntemann, C. Tamm, A. Surzhykov, and E. Peik, Phys. Rev. Lett. {\bf 127}, 213001 (2021).

\bibitem{Tofful-Metrologia-2024}
A. Tofful, C. F. A. Baynham, E. A. Curtis, A. O. Parsons, B. I. Robertson, M. Schioppo, J. Tunesi, H. S. Margolis, R. J. Hendricks, J. Whale, R. C. Thompson, and R. M. Godun, Metrologia {\bf 61}, 045001 (2024).

\bibitem{Aeppli2024}
A. Aeppli, K. Kim, W. Warfield, M. S. Safronova, and J. Ye, Phys. Rev. Lett. {\bf 113}, 023401 (2024).

\bibitem{Takamoto-NP-2020}
M. Takamoto, I. Ushijima, N. Ohmae, T. Yahagi, K. Kokado, H. Shinkai, and H. Katori, Nat. Photon. {\bf 14}, 411 (2020).

\bibitem{Kim2023}
K. Kim, A. Aeppli, T. Bothwell, and J. Ye, Phys. Rev. Lett. {\bf 130}, 113203 (2024).

\bibitem{Zhang2022}
A. Zhang, Z. X. Xiong, X. T. Chen, Y. Y. Jiang, J. Q. Wang, C. C. Tian, Q. Zhu, B. Wang, D. Z. Xiong, L. X. He, L. S. Ma, and B. L. Lyu, Metrologia {\bf 59}, 065009 (2022).

\bibitem{Luo2020}
L. M. Luo, H. Qiao, D. Ai, M. Zhou, S. Zhang, C. Y. Sun, Q. C. Qi, C. Q. Peng, T. Y. Jin, W. Fang, Z. Q. Yang, T. C. Li, K. Liang, and X. Y. Xu, Metrologia {\bf 57}, 065017 (2020).

\bibitem{Zeng2023}
M. Y. Zeng, Y. Huang, B. L. Zhang, Y. M. Hao, Z. X. Ma, R. M. Hu, H. Q. Zhang, Z. Chen, M. Wang, H. Guan, and K. L. Gao, Phys. Rev. Applied {\bf 19}, 064004 (2023).

\bibitem{Hausser-PRL-2025}
H. N. Hausser, J. Keller, T. Nordmann, N. M. Bhatt, J. Kiethe, H. Liu, I. M. Richter, M. von Boehn, J. Rahm, S. Weyers, E. Benkler, B. Lipphardt, S. D\"{o}rscher, K. Stahl, J. Klose, C. Lisdat, M. Filzinger, N. Huntemann, E. Peik, T. E. Mehlstäubler, Phys. Rev. Lett. {\bf 134}, 023201 (2025).

\bibitem{Flambaum2007}
V. V. Flambaum, Int. J. Mod. Phys. A {\bf 22}, 4937 (2007).

\bibitem{Godun2014}
R. M. Godun, P. B. R. Nisbet-Jones, J. M. Jones, S. A. King, L. A. M. Johnson, H. S. Margolis, K. Szymaniec, S. N. Lea, K. Bongs, and P. Gill, Phys. Rev. Lett. {\bf 113}, 210801 (2014).

\bibitem{Huntemann2014}
N. Huntemann, B. Lipphardt, C. Tamm, V. Gerginov, S. Weyers, and E. Peik, Phys. Rev. Lett. {\bf 113}, 210802 (2014).

\bibitem{Lange-PRL-2021a} 
R. Lange, N. Huntemann, J. M. Rahm, C. Sanner, H. Shao, B. Lipphardt, C. Tamm, S. Weyers, and E. Peik, Phys. Rev. Lett. {\bf 126}, 011102 (2021).

\bibitem{Filzinger2023}
M. Filzinger, S. D\"{o}rscher, R. Lange, J. Klose, M. Steinel, E. Benkler, E. Peik, C. Lisdat, and N. Huntemann, Phys. Rev. Lett. {\bf 130}, 253001 (2023).

\bibitem{Chou2010}	 
C. W. Chou, D. B. Hume, T. Rosenband, and D. J. Wineland, Science {\bf 329}, 1630 (2010).

\bibitem{Delva2018}
P. Delva, N. Puchades, E. Sch\"{o}nemann, F. Dilssner, C. Courde, S. Bertone, F. Gonzalez, A. Hees, C. L. Poncin-Lafitte, F. Meynadier, R. Prieto-Cerdeira, B. Sohet, J. Ventura-Traveset, and P. Wolf, Phys. Rev. Lett. {\bf 121}, 231101 (2018).

\bibitem{Bothwell-Nature-2022}
T. Bothwell, C. J. Kennedy, A. Aeppli, D. Kedar, J. M. Robinson, E. Oelker, A. Staron, and J. Ye,  Nature {\bf 602}, 420 (2022).

\bibitem{Megidish2019} 
E. Megidish, J. Broz, N. Greene, and H. H\"{a}ffner, Phys. Rev. Lett. {\bf 122}, 123605 (2019).

\bibitem{Sanner2019} 
C. Sanner, N. Huntemann, R. Lange, C. Tamm, E. Peik, M. S. Safronova, and S. G. Porsev, Nature {\bf 567}, 204 (2019).

\bibitem{Dreissen2022}
L. S. Dreissen, C. H. Yeh, H. A. F\"{u}rst, K. C. Grensemann, and T. E. Mehlst\"{o}ubler, Nat. Commun. {\bf 13}, 7314 (2022).

\bibitem{McGrew2018}
W. F. McGrew, X. Zhang, R. J. Fasano, S. A. Sch\"{a}ffer, K. Beloy, D. Nicolodi, R. C. Brown, N. Hinkley, G. Milani, M. Schioppo, T. H. Yoon, and A. D. Ludlow, Nature {\bf 564}, 87 (2018).

\bibitem{Beloy2021a}
K. Beloy, M. I. Bodine, T. Bothwell, S. M. Brewer, S. L. Bromley, J. S. Chen, J. D. Desch\^{e}nes, S. A. Diddams, R. J. Fasano, T. M. Fortier, Y. S. Hassan, D. B. Hume, D. Kedar, C. J. Kennedy, I. Khader, A. Koepke, D. R. Leibrandt, H. Leopardi, A. D. Ludlow, W. F. McGrew, W. R. Milner, N. R. Newbury, D. Nicolodi, E. Oelker, T. E. Parker, J. M. Robinson, S. Romisch, S. A. Sch\"{a}ffer, J. A. Sherman, L. C. Sinclair, L. Sonderhouse, W. C. Swann, J. Yao, J. Ye. and X. G. Zhang, Nature {\bf 591}, 564 (2021).

\bibitem{Ellis2021} 
J. L. Ellis, M. I. Bodine, W. C. Swann, S. A. Stevenson, E. D. Caldwell, L. C. Sinclair, N. R. Newbury, and J. D. Desch$\hat{e}$nes, Phys. Rev. Applied {\bf 15}, 034002 (2021).

\bibitem{Burt2021} 
E. A. Burt, J. D. Prestage, R. L. Tjoelker, D. G. Enzer, D. Kuang, D. W. Murphy, D. E. Robison, J. M. Seubert, R. T. Wang, and T. A. Ely, Nature {\bf 595}, 43 (2021).

\bibitem{Casini2023} 
S. Casini, E. Turan, A. Cervone, B. Monna, and P. Visser, Sci. Rep. {\bf 13}, 16253 (2023).

\bibitem{Burt2024}
E. A. Burt, T. A. Ely, and R. L. Tjoelker, J. Phys. Conf. Ser. {\bf 2889}, 012014 (2024).

\bibitem{Berengut-PRL-2018}
J. C. Berengut, D. Budker, C. Delaunay, V. V. Flambaum, C. Frugiuele, E. Fuchs, C. Grojean, R. Harnik, R. Ozeri, G. Perez, and Y. Soreq, Phys. Rev. Lett. {\bf 120}, 091801 (2018).

\bibitem{Berengut-PRR-2020}
J. C. Berengut, C. Delaunay, A. Geddes, and Y. Soreq, Phys. Rev. Research {\bf 2}, 043444 (2020).

\bibitem{Brzeminski-PRD-2022}
D. Brzeminski, Z. Chacko, A. Dev, I. Flood, and A. Hook, Phys. Rev. D {\bf 106}, 095031 (2022).

\bibitem{Rehbehn-PRL-2023}
N. H. Rehbehn, M. K. Rosner, J. C. Berengut, P. O. Schmidt, T. Pfeifer, M. F. Gu, and J. R. Crespo L\'{o}pez-Urrutia, Phys. Rev. Lett. {\bf 131}, 161803 (2023).

\bibitem{Dzuba-PRD-2024}
V. A. Dzuba, V. V. Flambaum, and A. J. Mansour, Phys. Rev. D {\bf 110}, 055022 (2024).

\bibitem{Berengut-NRP-2025}
J. C. Berengut and C. Delaunay, Nat. Rev. Phys. {\bf 7}, 119 (2025). 

\bibitem{Safronova2018}
M. S. Safronova, D. Budker, and D. DeMille, D. F. J. Kimball, A. Derevianko, and C. W. Clark, Rev. Mod. Phys. {\bf 90}, 025008 (2018).

\bibitem{Kozlov2018}
M. G. Kozlov, M. S. Safronova, J. R. Crespo L\'{o}pez-Urrutia, and P. O. Schmidt, Rev. Mod. Phys. {\bf 90}, 045005 (2018).

\bibitem{Yu-FP-2023}
Y. M. Yu, B. K. Sahoo, and B. B. Suo, Front. Phys. {\bf 11}, 1104848 (2023).

\bibitem{Berengut2011}
J. C. Berengut, V. A. Dzuba, V. V. Flambaum, and A. Ong, Phys. Rev. Lett. {\bf 106}, 210802 (2011).

\bibitem{Safronova2014a}
M. S. Safronova, V. A. Dzuba, V. V. Flambaum, U. I. Safronova, S. G. Porsev, and M. G. Kozlov, Phys. Rev. Lett. {\bf 113}, 030801 (2014).

\bibitem{Yudin2014}
V. I. Yudin, A. V. Taichenachev, and A. Derevianko, Phys. Rev. Lett. {\bf 113}, 233003 (2014).

\bibitem{Dzuba2015a}
V. A. Dzuba, V. V. Flambaum, and H. Katori, Phys. Rev. A {\bf 91}, 022119 (2015).

\bibitem{Dzuba2015b}
V. A. Dzuba and V. V. Flambaum, Hyperfine Interact. {\bf 79}, 86 (2015).

\bibitem{Dzuba2015c}
V. A. Dzuba, M. S. Safronova, U. I. Safronova, and V. V. Flambaum, Phys. Rev. A {\bf 92}, 060502 (2015).

\bibitem{Yu2016}
Y. M. Yu and B. K. Sahoo, Phys. Rev. A {\bf 94}, 062502 (2016).

\bibitem{Yu2018}
Y. M. Yu and B. K. Sahoo, Phys. Rev. A {\bf 97}, 041403 (2018).

\bibitem{Yu2019}
Y. M. Yu and B. K. Sahoo, Phys. Rev. A {\bf 99}, 022513 (2019).

\bibitem{Bekker2019}
H. Bekker, A. Borschevsky, Z. Harman, C. H. Keitel, T. Pfeifer, P. O. Schmidt, J. R. Crespo L\'{o}pez-Urrutia, and J. C. Berengut, Nat. Commun. {\bf 10}, 5651 (2019).

\bibitem{Beloy2021}
K. Beloy, Phys. Rev. Lett. {\bf 127}, 013201 (2021).

\bibitem{Yu2022}
Y. M. Yu, P. Duo, S. L. Chen, B. Arora, H. Guan, K. Gao, and J. Chen, Atoms {\bf 10}, 123 (2022).

\bibitem{Allehabi-PRA-2022}
S. O. Allehabi, S. M. Brewer, V. A. Dzuba, V. V. Flambaum, and K. Beloy, Phys. Rev. A {\bf 106}, 043101 (2022).

\bibitem{Porsev2024}
S. G. Porsev, C. Cheung, M. S. Safronova, H. Bekker, N. H. Rehbehn, J. R. Crespo L\'{o}pez-Urrutia, and S. M. Brewer, Phys. Rev. A {\bf 110}, 042823 (2024).

\bibitem{Dzuba2024a}
V. A. Dzuba and V. V. Flambaum, Phys. Rev. A {\bf 110}, 012801 (2024).

\bibitem{Yu-PRA-2024}
Y. M. Yu and B. K. Sahoo, Phys. Rev. A {\bf 109}, 023106 (2024).

\bibitem{Allehabi-JQSRT-2024}
S. O. Allehabi, V. A. Dzuba, and V. V. Flambaum, J. Quant. Spectr. Rad. Transfer {\bf 328}, 109151 (2004).

\bibitem{King2022}
S. A. King, L. J. Spie\ss, P. Micke, A. Wilzewski, T. Leopold, E. Benkler, R. Lange, N. Huntemann, A. Surzhykov, V. A. Yerokhin, J. R. Crespo L\'{o}pez-Urrutia, and P. O. Schmidt, Nature {\bf 611}, 43 (2022).

\bibitem{Berengut2010}
J. C. Berengut, V. A. Dzuba, and V. V. Flambaum, Phys. Rev. Lett. {\bf 105}, 120801 (2010).

\bibitem{Windberger2015}
A. Windberger, J. R. Crespo L\'{o}pez-Urrutia, H. Bekker, N. S. Oreshkina, J. C. Berengut, V. Bock, A. Borschevsky, V. A. Dzuba, E. Eliav, Z. Harman, U. Kaldor, S. Kaul, U. I. Safronova, V. V. Flambaum, C. H. Keitel, P. O. Schmidt, J. Ullrich, and O. O. Versolato, Phys. Rev. Lett. {\bf 114}, 150801 (2015).

\bibitem{Safronova2015}
U. I. Safronova, V. V. Flambaum, and M. S. Safronova, Phys. Rev. A {\bf 92}, 022501 (2015).

\bibitem{Cheung2020}
C. Cheung, M. S. Safronova, S. G. Porsev, M. G. Kozlov, I. I. Tupitsyn, and A. I. Bondarev, Phys. Rev. Lett. {\bf 124}, 163001 (2020).

\bibitem{Dzuba2023}
V. A. Dzuba and V. V. Flambaum, Phys. Rev. A {\bf 108}, 053111 (2023).

\bibitem{Olsen1990}
J. Olsen, P. J{\o}rgensen, and J. Simons, Chem. Phys. Lett. {\bf 169}, 463 (1990).

\bibitem{Olsen-JCP-2000}
J. Olsen, J. Chem. Phys. {\bf 113}, 7140 (2000).

\bibitem{Fleig2003}
T. Fleig, J. Olsen, and L. Visscher, J. Chem. Phys. {\bf 119}, 2963 (2003).

\bibitem{Fleig2006}
T. Fleig, H. J. Aa. Jensen, J. Olsen, and L. Visscher, J. Chem. Phys. {\bf 124}, 104106 (2006). 

\bibitem{Thyssen-JCP-2008}
J. Thyssen, T. Fleig, and H. J. Aa. Jensen, J. Chem. Phys. {\bf 129}, 034109 (2008).

\bibitem{Knecht-JCP-2010}
S. Knecht, H. J. Aa. Jensen, and T. Fleig, J. Chem. Phys. {\bf 132}, 014108 (2010).

\bibitem{Kaldor1991}
U. Kaldor, Theor. Chim. Acta {\bf 80}, 427 (1991).

\bibitem{Visscher-JCP-1996}
L. Visscher, T. J. Lee and K. G. Dyall, J. Chem. Phys. {\bf 105} (1996) 8769.

\bibitem{Visscher-JCP-2002}
L. Visscher, E. Eliav and U. Kaldor, J. Chem. Phys. {\bf 115} (2002) 9720.

\bibitem{Pernpointner-JCP-2003}
M. Pernpointner and L. Visscher, J. Comp. Chem. {\bf 24} 754 (2003).

\bibitem{Oleynichenko2020}
A. V. Oleynichenko, A. Zaitsevskii, and E. Eliav, in Super-computing, edited by V. Voevodin and S. Sobolev (Springer, Cham, 2020), Vol. {\bf 1331}, pp. 375–386.

\bibitem{Oleynichenko2020a}
A. Oleynichenko, A. Zaitsevskii, and E. Eliav, EXP-T, an extensible code for Fock space relativistic coupled cluster calculations (see http://www.qchem.pnpi.spb.ru/expt) (2020).

\bibitem{Dyall2023}
K. G. Dyall, P. Tecmer, and A. Sunaga, J. Chem. Theory Comput. {\bf 19}, 198 (2023).

\bibitem{Visscher-ADNDT-1997}
L. Visscher and K. G. Dyall, At. Data Nucl. Data Tables {\bf 67}, 207 (1997).

\bibitem{Sikkema-JCP-2009}
J. Sikkema, L. Visscher, T. Saue and M. Ilia\v{s}, J. Chem. Phys. {\bf 131}, 124116 (2009).

\bibitem{Knecht-JCP-2022}
S. Knecht, M. Repisky, H. J. Aa. Jensen, and T. Saue, J. Chem. Phys. {\bf 157}, 114106 (2022).

\bibitem{Fleig2001}
T. Fleig, J. Olsen, and C. M. Marian, J. Chem. Phys. {\bf 114}, 4775 (2001).

\bibitem{Thyssen-AOC}
J. Thyssen, Ph.D. dissertation, Department of Chemistry, University of Southern Denmark, Odense, Denmark, 2001.

\bibitem{SM}
Detailed investigation of DHF reference states, data of the KRCI and FSCC calculations of energy and various properties for different basis sets and computational parameters with comparison of data and uncertainty analysis, the calculation of the lifetime of excited states. 

\bibitem{Dirac}
DIRAC, a relativistic ab initio electronic structure program, Release DIRAC22 (2022), written by H. J. {\relax Aa. Jensen}, R. Bast, A. S. P. Gomes, T. Saue and L. Visscher, with contributions from I. A. Aucar, V. Bakken, C. Chibueze, J. Creutzberg, K. G. Dyall,
S. Dubillard, U. Ekstr{\"o}m, E. Eliav, T. Enevoldsen, E. Fa{\ss}hauer, T. Fleig, O. Fossgaard, L. Halbert, E. D. Hedeg{\aa}rd, T. Helgaker, B. Helmich--Paris, J. Henriksson, M. van Horn, M. Ilia{\v{s}}, Ch. R. Jacob, S. Knecht, S. Komorovsk{\'y}, O. Kullie, J. K. L{\ae}rdahl, C. V. Larsen, Y. S. Lee, N. H. List, H. S. Nataraj, M. K. Nayak, P. Norman, G. Olejniczak, J. Olsen, J. M. H. Olsen, A. Papadopoulos, Y. C. Park, J. K. Pedersen, M. Pernpointner, J. V. Pototschnig, R. Di Remigio, M. Repisky, K. Ruud, P. Sa{\l}ek, B. Schimmelpfennig, B. Senjean, A. Shee, J. Sikkema, A. Sunaga, A. J. Thorvaldsen, J. Thyssen, J. van Stralen, M. L. Vidal, S. Villaume, O. Visser, T. Winther, S. Yamamoto and X. Yuan (available at \url{http://dx.doi.org/10.5281/zenodo.6010450}, see also  \url{http://www.diracprogram.org}).

\bibitem{Dirac2020}
T. Saue, R. Bast, A. S. P. Gomes, H. J. Aa. Jensen, L. Visscher, I. A. Aucar, R. Di Remigio, K. G. Dyall, E. Eliav, E. Fa{\ss}hauer, T. Fleig, L. Halbert, E. D. Hedeg{\aa}rd, B. Helmich-Paris, M. Ilia\v{s}, C. R. Jacob, S. Knecht, J. K. Laerdahl, M. L. Vidal, M. K. Nayak, M. Olejniczak, J. M. H. Olsen, M. Pernpointner, B. Senjean, A. Shee, A. Sunaga, and J. N. P. van Stralen, J. Chem. Phys. {\bf 152}, 204104 (2020).

\bibitem{Peng2013}
D. Peng, N. Middendorf, F. Weigend, M. Reiher, J. Chem. Phys. {\bf 138}, 184105 (2013).

\bibitem{Archibong1991}
E. F. Archibong and A. J. Thakkar, Phys. Rev. A {\bf 44}, 5478 (1991).

\bibitem{Guo-PRA-2021}
X. T. Guo, Y. M. Yu, Y. Liu, B. B. Suo, and B. K. Sahoo, Phys. Rev. A {\bf 103}, 013109 (2021). 

\bibitem{Haase-JPCA-2020}
P. A. B. Haase, E. Eliav, M. Ilia\'{s}, and A. Borschevsky, J. Phys. Chem. A {\bf 124}, 3157 (2020). 

\bibitem{Denis-PRA-2022}
M. Denis, P. A. B. Haase, M. C. Mooij, Phys. Rev. A {\bf 105}, 052811 (2022). 

\bibitem{Dube-PRA-2013}
P. Dub\'{e}, A. A. Madej, Z. Zhou, and J. E. Bernard, Phys. Rev. A {\bf 87}, 023806 (2013). 

\bibitem{Shaniv2018}
R. Shaniv, N. Akerman, T. Manovitz, Y. Shapira, N. Ozeri, and D. F. J. Kimball, Phys. Rev. Lett. {\bf 120}, 103202 (2018).

\end{thebibliography}
\end{document}


\section{Choice of reference state for DHF calculation}	

In this section, we conduct a detailed investigation into the selection of the Dirac-Hartree-Fock (DHF) reference state, with the primary aim of preparing an appropriate initial set of Dirac spinors for subsequent Kramers-restricted configuration-interaction (KRCI) calculations. The Ir$^{17+}$ ion serves as a prototypical open-shell system, characterized by a relatively large number of outer-shell electrons. It contains 14 valence electrons, which can occupy both the $4f$ and $5s$ shells, leading to several possible ground-state configurations. Our research focuses on a two-fold investigation. First, we evaluate the impact of different initial reference-state specifications on the outcomes of KRCI calculations. Second, we identify the most suitable initial reference state for KRCI computations. Additionally, we quantify the uncertainties that may arise due to variations in reference-state choices. To this end, we examine four distinct cases: Case I: The 14 valence electrons are uniformly distributed across 8 spinor pairs (comprising 7 $4f$ and 1 $5s$), denoted as “14in8(4f5s)”; Case II: Since the $4f$ spinors have lower energy than the $5s$ spinors, all 14 valence electrons are placed in the 7 $4f$ spinor pairs, labeled as “14in7(4f)”; Case III: 13 valence electrons are distributed among the 7 $4f$ spinor pairs, while the remaining 1 valence electron occupies the $5s$ spinor, recorded as “13in7(4f)1in1(5s)”; Case IV: 12 valence electrons occupy the 7 $4f$ spinor pairs, and the remaining 2 valence electrons are assigned to a single $5s$ spinor pair, denoted as “12in7(4f)2in1(5s)”.

We employ the e24-CISD model, based on DHF calculations, to perform KRCI calculations for the four cases described above. The results are presented in Table \ref{tab:Ir17+-scfI}. A detailed analysis of these results reveals that, under the four different configurations, the excitation energies (EE) of the excited states for the Ir$^{17+}$ ion calculated using the “14in8(4f5s)” and “14in7(4f)” configurations tend to be higher, while those derived from the “13in7(4f)1in1(5s)” and “12in7(4f)2in1(5s)” configurations fall within a lower energy range. This energy differential is particularly pronounced for the $4f~^1S_0$ state. Comparing these results with previously reported computational findings in Refs. \cite{Cheung2020} and \cite{Dzuba2023}, it becomes evident that the results from the “13in7(4f)1in1(5s)” and “12in7(4f)2in1(5s)” configurations are more consistent with the expected physical behavior. However, it remains challenging to determine which of the two configurations, “13in7(4f)1in1(5s)” or “12in7(4f)2in1(5s),” is more advantageous.

In light of this issue, we extend our research methodology by employing the multi-configuration self-consistent field (MCSCF) method to optimize the wave functions derived from the four DHF spinors. The optimizations are performed using the “13in7(4f)1in1(5s)” and “14in7(4f)” complete active space (CAS) models, respectively. The KRCI results based on the MCSCF wave function for these two CAS models are presented in Tables \ref{tab:Ir17+-scfII} and \ref{tab:Ir17+-scfIII}, respectively. As shown in Table \ref{tab:Ir17+-scfII}, under the MCSCF optimization with the “13in7(4f)1in1(5s)” CAS model, the KRCI algorithm fails to accurately reproduce the $4f~^1S_0$ state, and the ground state is erroneously predicted as the $4f^{12}5s^{2}$ configuration. In contrast, for the “14in7(4f)” CAS model, despite starting from four distinct DHF calculations, the final results demonstrate a high degree of consistency with theoretical expectations. Furthermore, whether or not an additional MCSCF optimization step is included, the variation in the KRCI results remains negligible when the DHF calculations are initiated from the “14in7(4f)” reference state.

Based on this comprehensive and in-depth analysis, we ultimately select the “14in7(4f)” reference state for the DHF calculations in subsequent KRCI computations.
  
\begin{table}[htp]
\centering		
\caption{The excitation energies (EE) (in cm$^{-1}$) of the ground and low-lying excited states of the Ir$^{17+}$ ion are obtained using e24-CISD calculations based on Dirac spinors. These calculations are performed starting from four different AOC open-shell DHF reference states: Case I (14in8(4f5s)), Case II (14in7(4f)), Case III (13in7(4f)1in1(5s)), and Case IV (12in7(4f)2in1(5s)).\label{tab:Ir17+-scfI}}{\setlength{\tabcolsep}{8pt}
\begin{tabular}{ccc ccc}\hline\hline  \addlinespace[0.1cm]
Config.	&Term &	14in8(4f5s)	&	14in7(4f)	&	13in7(4f)1in1(5s)	&	12in7(4f)2in1(5s)	\\\hline \addlinespace[0.1cm]
$4f^{13}5s$	&	$^3F_4^o$	&	0 	&	0 	&	0 	&	0 	\\
&	$^3F_3^o$	&	4860 	&	4816 	&	4810 	&	4795 	\\
&	$^3F_2^o$	&	26046 	&	26188 	&	26082 	&	25973 	\\
&	$^1F_3^o$	&	31538 	&	31498 	&	31391 	&	31266 	\\ \addlinespace[0.1cm]
$4f^{14}$	&	$^1S_0$	&	10005 	&	11194 	&	15403 	&	20199 	\\ \addlinespace[0.1cm]
$4f^{12}5s^2$	&	$^3H_6$	&	26453 	&	28594 	&	24985 	&	20879 	\\
&	$^3F_4$	&	36195 	&	38523 	&	34914 	&	30820 	\\
&	$^3H_5$	&	50686 	&	52944 	&	49249 	&	45044 	\\
&	$^3F_2$	&	59177 	&	61399 	&	57760 	&	53642 	\\
&	$^1G_4$	&	59703 	&	62082 	&	58377 	&	54169 	\\
&	$^3F_3$	&	62647 	&	64928 	&	61232 	&	57038 	\\
&	$^3H_4$	&	82852 	&	85402 	&	81615 	&	77316 	\\
&	$^1D_2$	&	88362 	&	90367 	&	86685 	&	82537 	\\
&	$^1J_6$	&	100537 	&	101335 	&	97653 	&	93493 	\\ \hline \hline  
	\end{tabular}}
\end{table}	

\begin{table}[htp]
	\centering		
	\caption{The excitation energies (EE) (in cm$^{-1}$) of the ground and low-lying excited states of the Ir$^{17+}$ ion are calculated using e24-CISD based on the MCSCF wave function under the 14in8(4f5s) CAS. These calculations are performed for the 14in8(4f5s), 14in7(4f), 13in7(4f)1in1(5s), and 12in7(4f)2in1(5s) DHF reference states. \label{tab:Ir17+-scfII}}{\setlength{\tabcolsep}{8pt}
		\begin{tabular}{ccc ccc}\hline\hline  \addlinespace[0.1cm]
			Config.	&Term &		14in8(4f5s)	&	14in7(4f)	&	13in7(4f)1in1(5s)	&	12in7(4f)2in1(5s)	\\\hline \addlinespace[0.1cm]
			$4f^{13}5s$	&	$^3F_4^o$	&	231346 	&	0	&	0	&	237384 	\\
			&	$^3F_3^o$	&	238872 	&	4857 	&	4858 	&	243052 	\\
			&	$^3F_2^o$	&	260647 	&	26090 	&	26090 	&	263644 	\\
			&	$^1F_3^o$	&	265773 	&	31479 	&	31480 	&	270556 	\\ \addlinespace[0.1cm]
			$4f^{14}$	&	$^1S_0$	&		&	11464 	&	11464 	&		\\ \addlinespace[0.1cm]
			$4f^{12}5s^2$	&	$^3H_6$	&	0 	&	27329 	&	27330 	&	0 	\\
			&	$^3F_4$	&	11239 	&	37255 	&	37255 	&	10239 	\\
			&	$^3H_5$	&	25459 	&	51589 	&	51589 	&	24515 	\\
			&	$^3F_2$	&	34142 	&	60091 	&	60091 	&	33144 	\\
			&	$^1G_4$	&	34651 	&	60715 	&	60715 	&	33709 	\\
			&	$^3F_3$	&	37664 	&	63566 	&	63567 	&	36407 	\\
			&	$^3H_4$	&	59031 	&	83949 	&	83949 	&	57582 	\\
			&	$^1D_2$	&	63055 	&	88996 	&	88995 	&	61904 	\\
			&	$^1J_6$	&	73178 	&	99950 	&	99948 	&	72647 	\\ \hline \hline  
	\end{tabular}}
\end{table}

\begin{table}[htp]
	\centering		
	\caption{The EE (in cm$^{-1}$) of the ground and low-lying excited states of the Ir$^{17+}$ ion are obtained using e24-CISD calculations based on the MCSCF wave function under the 14in8(4f5s) CAS. These calculations are performed for the DHF reference states: 14in8(4f5s), 14in7(4f), 13in7(4f)1in1(5s), and 12in7(4f)2in1(5s).  \label{tab:Ir17+-scfIII}}{\setlength{\tabcolsep}{8pt}
		\begin{tabular}{ccc ccc}\hline\hline  \addlinespace[0.1cm]
Config.	&Term 	&	14in8(4f5s)	&	14in7(4f)	&	13in7(4f)1in1(5s)	&	12in7(4f)2in1(5s)	\\\hline \addlinespace[0.1cm]
$4f^{13}5s$	&	$^3F_4^o$	&	0 	&	0 	&	0 	&	0 	\\
&	$^3F_3^o$	&	4830 	&	4814 	&	4825 	&	4830 	\\
&	$^3F_2^o$	&	26103 	&	26106 	&	26104 	&	26103 	\\
&	$^1F_3^o$	&	31443 	&	31420 	&	31437 	&	31444 	\\ \addlinespace[0.1cm]
$4f^{14}$	&	$^1S_0$	&	12111 	&	11840 	&	12078 	&	12111 	\\ \addlinespace[0.1cm]
$4f^{12}5s^2$	&	$^3H_6$	&	27482 	&	28032 	&	27606 	&	27476 	\\
&	$^3F_4$	&	37403 	&	37957 	&	37528 	&	37397 	\\
&	$^3H_5$	&	51757 	&	52310 	&	51882 	&	51751 	\\
&	$^3F_2$	&	60239 	&	60800 	&	60366 	&	60233 	\\
&	$^1G_4$	&	60881 	&	61438 	&	61007 	&	60875 	\\
&	$^3F_3$	&	63730 	&	64287 	&	63856 	&	63724 	\\
&	$^3H_4$	&	84129 	&	84690 	&	84257 	&	84123 	\\
&	$^1D_2$	&	89148 	&	89713 	&	89276 	&	89142 	\\
&	$^1J_6$	&	100108 	&	100672 	&	100236 	&	100101 	\\ \hline \hline  
	\end{tabular}}
\end{table}

\clearpage

\section{Analysis of final values and uncertainty for KRCI and FSCC calculations}	
In this section, we present a detailed analysis of the KRCI and FSCC calculations performed using various computational parameters. These parameters include the size of the basis set, the truncation of virtual orbitals, and the ranks of higher-order excitations. The results are summarized for the excitation energies (shown in Tables \ref{tab:Ir17+_EE_KRCI} and \ref{tab:Ir17+_EE_EXPT}), the electric dipole polarizabilities (presented in Tables \ref{tab:Ir17+_alpha_krci} and \ref{tab:Ir17+_alpha_DIRAC}), the electric quadrupole moments (displayed in Tables \ref{tab:Ir17+_theta_krci}, \ref{tab:Ir17+_theta_DIRAC}, and \ref{tab:Ir17+_theta_EXPT}), the hyperfine constants (recorded in Tables \ref{tab:Ir17+_HFSA-KRCI} and \ref{tab:Ir17+_HFSA_DIRAC}), and the field shifts (illustrated in Tables \ref{tab:Ir17+_FS_DIRAC} and \ref{tab:Ir17+_FS_EXPT}). These detailed tabular data provide a robust foundation for presenting our final results and associated uncertainties. For the FSCC calculations, we employ both the DIRAC and EXPT codes whenever possible to ensure cross-verification.
   
\subsection{Energy}	

\begin{table}[htp]
	\centering		
\caption{The EE (in cm$^{-1}$) of the ground state and low-lying excited states of the Ir$^{17+}$ ion is calculated using the KRCI method. The notations `e14', `e24’, and `e32' represent the number of active electrons in the configuration interaction (CI) model. The e32-CISD$<20$ calculations are performed for the dyall.aae2z ($2\xi$), dyall.aae3z ($3\xi$), and dyall.aae4z ($4\xi$) basis sets to obtain size-converged results for the basis sets, where $<20$ indicates the truncation of virtual orbitals with energy less than 20 a.u. We define $\Delta_{basis}$ as the impact of the finite size of the basis set, estimated by the difference between the e32-CISD$<20$ values for the $3\xi$ and $4\xi$ basis sets. The correction due to the truncation of virtual orbitals is represented by $\Delta_{virt}$, estimated as the difference between e24-CISD calculations with truncations of virtual orbitals less than 20 a.u. ($<20$) and 300 a.u. ($<300$) under the dyall.aae2z basis set. We use $\Delta_{T}$ to quantify the influence of triple excitations of the 14 valence electrons, estimated as the difference between e14-CI $2\xi<20$ values for single-double (SD) and single-double-triple (SDT) excitations under the $2\xi<20$ condition. The e32-CISD$<20$ values under the $4\xi$ basis set are regarded as the final results, denoted as `FINAL', and the uncertainty (`Uncert.') is evaluated using the formula $\sqrt{\Delta_{T}^2 + \Delta_{virt}^2 + \Delta_{basis}^2}$, which is quoted in the main text under `e32-CISD' of Table I. \label{tab:Ir17+_EE_KRCI}}{\setlength{\tabcolsep}{4pt}
	    \begin{tabular}{ccc ccc ccc ccc ccc cc}\hline\hline  \addlinespace[0.1cm]
\multirow{2}{*}{Config.}	&	\multirow{2}{*}{Term}	&&	\multicolumn{3}{c}{e14-CI $2\xi<20$}					&&	\multicolumn{3}{c}{e24-CISD $2\xi$}					&&	\multicolumn{4}{c}{e32-CISD$<20$}							&	\multirow{2}{*}{Uncert.}	\\ \cline{4-6} \cline{8-10}  \cline{12-15} \addlinespace[0.1cm]
&		&&	SD 	&	SDT	&	$\Delta_T$	&&	$<20$	&	$<300$	&	$\Delta_{virt}$	&&	$2\xi$	&	$3\xi$	&	$4\xi$ FINAL	&	$\Delta_{basis}$	&		\\ \hline \addlinespace[0.1cm]
(1)	&	(2)	&&	(3)	&	(4)	&	(5)	&&	(6)	&	(7)	&	(8)	&&	(9)	&	(10)	&	(11)	&	(12)	&	(13)	\\ \addlinespace[0.1cm]
$4f^{13}5s$	&	$^3F_4^o$	&&	0	&	0	&	0 	&&	0 	&	0 	&	0 	&&	0 	&	0 	&	0 	&	0 	&	0 	\\
&	$^3F_3^o$	&&	4820 	&	4822 	&	2 	&&	4816 	&	4807 	&	-9 	&&	4847 	&	4813 	&	4769 	&	-44 	&	45 	\\
&	$^3F_2^o$	&&	26037 	&	26047 	&	10 	&&	26188 	&	26178 	&	-10 	&&	26126 	&	26216 	&	26194 	&	-22 	&	26 	\\
&	$^1F_3^o$	&&	31253 	&	31265 	&	12 	&&	31498 	&	31483 	&	-15 	&&	31482 	&	31488 	&	31381 	&	-107 	&	109 	\\
$4f^{14}$	&	$^1S_0$	&&	14230 	&	12934 	&	-1296 	&&	11194 	&	11593 	&	399 	&&	10778 	&	12322 	&	12006 	&	-316 	&	1392 	\\
$4f^{12}5s^2$	&	$^3H_6$	&&	27242 	&	27057 	&	-184 	&&	28594 	&	28797 	&	203 	&&	27754 	&	26749 	&	28848 	&	2099 	&	2117 	\\
&	$^3F_4$	&&	37023 	&	36841 	&	-182 	&&	38523 	&	38694 	&	171 	&&	37214 	&	36217 	&	38221 	&	2003 	&	2019 	\\
&	$^3H_5$	&&	51498 	&	51300 	&	-198 	&&	52944 	&	53147 	&	203 	&&	52004 	&	51067 	&	53162 	&	2096 	&	2115 	\\
&	$^3F_2$	&&	60101 	&	59901 	&	-200 	&&	61399 	&	61548 	&	149 	&&	59916 	&	59042 	&	60530 	&	1489 	&	1509 	\\
&	$^1G_4$	&&	60535 	&	60338 	&	-198 	&&	62082 	&	62264 	&	181 	&&	60843 	&	59913 	&	61953 	&	2040 	&	2058 	\\
&	$^3F_3$	&&	63550 	&	63345 	&	-205 	&&	64928 	&	65114 	&	186 	&&	63668 	&	62803 	&	64577 	&	1774 	&	1795 	\\
&	$^3H_4$	&&	83678 	&	83472 	&	-206 	&&	85402 	&	85572 	&	170 	&&	83984 	&	83084 	&	85218 	&	2134 	&	2150 	\\
&	$^1D_2$	&&	89364 	&	89141 	&	-223 	&&	90367 	&	90508 	&	142 	&&	88951 	&	88226 	&	89257 	&	1031 	&	1064 	\\
&	$^1J_6$	&&	101282 	&	101075 	&	-207 	&&	101335 	&	101482 	&	147 	&&	100169 	&	99481 	&	100121 	&	640 	&	688 	\\\hline\hline
		\end{tabular}}
	\end{table}

\begin{table}[H]
	\centering		
\caption{The EE (in cm$^{-1}$) of the Ir$^{17+}$ ion is calculated using the FSCC method implemented within the EXPT code. The e60-CCSD calculations are performed with the dyall.aae2z, dyall.aae3z, and dyall.aae4z basis sets. The notations `e32' and `e60' represent the number of correlated electrons in the coupled-cluster (CC) calculations. The data $\Delta_T$ accounts for the influence of triple-rank excitations in the CC amplitudes and is estimated as the difference between the values obtained from e32-CC calculations under single-double (SD) and single-double-triple (SDT) excitations with the $2\xi$ basis set. The data $\Delta_{basis}$ considers the effect of the finite size of the basis set and is estimated as the difference between e60-CCSD values obtained with the $3\xi$ and $4\xi$ basis sets. Adding the corrections $\Delta_T$ and $\Delta_{basis}$ to the e60-CCSD results with the $4\xi$ basis set yields the final values (denoted as `FINAL'), while the uncertainty (`Uncert.') is calculated as $\sqrt{ {\Delta_{T}}^2 + {\Delta_{basis}}^2 }$, which is quoted in Table II under `e60-CCSD' of the main text. \label{tab:Ir17+_EE_EXPT}}{\setlength{\tabcolsep}{9pt}
	    \begin{tabular}{ccc ccc ccc ccc c}\hline\hline  \addlinespace[0.1cm]
\multirow{2}{*}{Config.}	&	\multirow{2}{*}{Term}	&&	\multicolumn{3}{c}{e32-CC $2\xi<20$}					&&	\multicolumn{4}{c}{e60-CCSD}							&	\multirow{2}{*}{FINAL}	&	\multirow{2}{*}{Uncert.}	\\ \cline{4-6}  \cline{8-11} \addlinespace[0.1cm]
&		&&	SD	&	SDT	&	$\Delta_{T}$	&&	$2\xi$	&	$3\xi$	&	$4\xi$	&	$\Delta_{basis}$	&		&		\\ \hline \addlinespace[0.1cm]
(1)	&	(2)	&&	(3)	&	(4)	&	(5)	&&	(6)	&	(7)	&	(8)	&	(9)	&	(10)	&	(11)	\\ \addlinespace[0.1cm]
$4f^{13}5s$	&	$^3F_4^o$	&&	0 	&	0 	&	0 	&&	0 	&	0 	&	0 	&	0 	&	0 	&	0 	\\
&	$^3F_3^o$	&&	4757 	&	4814 	&	57 	&&	4753 	&	4693 	&	4646 	&	-48 	&	4655 	&	74 	\\
&	$^3F_2^o$	&&	25106 	&	25232 	&	126 	&&	25187 	&	25291 	&	25321 	&	30 	&	25476 	&	129 	\\
&	$^1F_3^o$	&&	30431 	&	30657 	&	225 	&&	30492 	&	30462 	&	30408 	&	-54 	&	30579 	&	232 	\\
$4f^{14}$	&	$^1S_0$	&&	13882 	&	10441 	&	-3441 	&&	15140 	&	15490 	&	14567 	&	-923 	&	10203 	&	3563 	\\
$4f^{12}5s^2$	&	$^3H_6$	&&	22081 	&	24810 	&	2729 	&&	21039 	&	22194 	&	23455 	&	1261 	&	27445 	&	3007 	\\
&	$^3F_4$	&&	31204 	&	33693 	&	2489 	&&	30022 	&	31124 	&	32348 	&	1225 	&	36062 	&	2774 	\\
&	$^3H_5$	&&	45326 	&	48164 	&	2839 	&&	44367 	&	45633 	&	46926 	&	1294 	&	51059 	&	3120 	\\
&	$^3F_2$	&&	53101 	&	55344 	&	2242 	&&	51695 	&	52376 	&	53532 	&	1157 	&	56931 	&	2523 	\\
&	$^1G_4$	&&	53867 	&	56558 	&	2691 	&&	52834 	&	54083 	&	55355 	&	1272 	&	59318 	&	2977 	\\
&	$^3F_3$	&&	56571 	&	59181 	&	2609 	&&	55419 	&	56416 	&	57650 	&	1234 	&	61492 	&	2886 	\\
&	$^3H_4$	&&	76122 	&	78897 	&	2774 	&&	75188 	&	76632 	&	77939 	&	1306 	&	82019 	&	3066 	\\
&	$^1D_2$	&&	81343 	&	83426 	&	2083 	&&	79818 	&	80150 	&	81241 	&	1091 	&	84415 	&	2351 	\\
&	$^1J_6$	&&	94372 	&	96317 	&	1946 	&&	92975 	&	92701 	&	93543 	&	842 	&	96331 	&	2120 	\\\hline\hline
	\end{tabular}}
	\end{table}

\subsection{$\alpha_d^{S,T}$}	

\begin{table}[H]
\centering		
\caption{The electric dipole polarizabilities $\alpha_d^{S,T}$ (in atomic units, a.u.) of the Ir$^{17+}$ ion are calculated using the KRCI finite-field (FF) approach. The e32-configuration interaction with single and double excitations (CISD) calculations are performed with the dyall.aae2z ($2\xi$) and dyall.aae3z ($3\xi$), considering the truncation of virtual orbitals with energies less than 20 atomic units ($<20$). The uncertainty, denoted as `Uncert.', primarily reflects the influence of the finite size of the basis set and is estimated as the difference between the values obtained with the $2\xi$ and $3\xi$ basis sets. The values in the columns labeled `FINAL' and `Uncert.' are reported in Table III of the main text as the results of the e32-CISD calculations for the electric dipole polarizabilities $\alpha_d^{S,T}$. \label{tab:Ir17+_alpha_krci}}{\setlength{\tabcolsep}{6pt}
	    \begin{tabular}{cccc cccc cccc cc}\hline\hline  \addlinespace[0.1cm]
\multirow{2}{*}{Config.}	&	\multirow{2}{*}{Term}	&&	\multicolumn{3}{c}{$2\xi$}					&&	\multicolumn{3}{c}{$3\xi$ FINAL}					&&	\multicolumn{3}{c}{Uncert.}					\\ \cline{4-6}  \cline{8-10} \addlinespace[0.1cm]
&		&&	$\alpha^S$	&	$\alpha^T$	&	$\Delta \alpha^S$	&&	$\alpha^S$	&	$\alpha^T$	&	$\Delta \alpha^S$	&&	$\alpha^S$	&	$\alpha^T$	&	$\Delta \alpha^S$	\\\hline \addlinespace[0.1cm]
$4f^{13}5s$	&	$^3F_4^o$	&&	0.3373 	&	0.0097 	&	0.0000 	&&	0.3600 	&	0.0085 	&	0.0000 	&&	0.0227 	&	-0.0012 	&	0.0000 	\\
&	$^3F_3^o$	&&	0.3389 	&	0.0066 	&	0.0016 	&&	0.3607 	&	0.0046 	&	0.0007 	&&	0.0217 	&	-0.0020 	&	-0.0009 	\\
&	$^3F_2^o$	&&	0.3371 	&	0.0064 	&	-0.0002 	&&	0.3591 	&	0.0043 	&	-0.0009 	&&	0.0220 	&	-0.0021 	&	-0.0007 	\\
&	$^1F_3^o$	&&	0.3379 	&	0.0037 	&	0.0005 	&&	0.3604 	&	0.0025 	&	0.0004 	&&	0.0226 	&	-0.0012 	&	-0.0001 	\\ \addlinespace[0.1cm]
$4f^{14}$	&	$^1S_0$	&&	0.1104 	&	0.0000 	&	-0.2269 	&&	0.1570 	&	0.0000 	&	-0.2030 	&&	0.0466 	&	0.0000 	&	0.0239 	\\ \addlinespace[0.1cm]
$4f^{12}5s^2$	&	$^3H_6$	&&	0.4723 	&	0.0109 	&	0.1350 	&&	0.4850 	&	0.0106 	&	0.1250 	&&	0.0127 	&	-0.0004 	&	-0.0100 	\\
&	$^3F_4$	&&	0.4789 	&	0.0055 	&	0.1416 	&&	0.4853 	&	-0.0011 	&	0.1253 	&&	0.0064 	&	-0.0066 	&	-0.0162 	\\
&	$^3H_5$	&&	0.4716 	&	0.0099 	&	0.1343 	&&	0.4842 	&	0.0101 	&	0.1242 	&&	0.0126 	&	0.0003 	&	-0.0101 	\\
&	$^3F_2$	&&	0.4720 	&	-0.0040 	&	0.1346 	&&	0.4846 	&	-0.0059 	&	0.1246 	&&	0.0126 	&	-0.0019 	&	-0.0100 	\\
&	$^1G_4$	&&	0.4707 	&	0.0044 	&	0.1334 	&&	0.4845 	&	0.0042 	&	0.1245 	&&	0.0137 	&	-0.0003 	&	-0.0090 	\\
&	$^3F_3$	&&	0.4711 	&	-0.0035 	&	0.1337 	&&	0.4845 	&	-0.0021 	&	0.1245 	&&	0.0135 	&	0.0014 	&	-0.0092 	\\
&	$^3H_4$	&&	0.4704 	&	0.0065 	&	0.1331 	&&	0.4835 	&	0.0048 	&	0.1235 	&&	0.0131 	&	-0.0018 	&	-0.0096 	\\
&	$^1D_2$	&&	0.4714 	&	-0.0037 	&	0.1341 	&&	0.4839 	&	-0.0033 	&	0.1239 	&&	0.0125 	&	0.0004 	&	-0.0102 	\\
&	$^1J_6$	&&	0.4706 	&	0.0222 	&	0.1333 	&&	0.4842 	&	0.0203 	&	0.1242 	&&	0.0136 	&	-0.0020 	&	-0.0090 	\\ \hline \hline  
	\end{tabular}}
	\end{table}
	
\begin{table}[H]
\centering		
\caption{The electric dipole polarizabilities $\alpha_d^{S,T}$ (in atomic units, a.u.) of the Ir$^{17+}$ ion are calculated using the FSCC finite-field (FF) approach implemented in the DIRAC code. The e60-CCSD calculations are performed with the dyall.aae2z ($2\xi$), dyall.aae3z ($3\xi$), and dyall.aae4z ($4\xi$) basis sets. The results obtained with the $4\xi$ basis set are considered as the final values, denoted as `FINAL.' The uncertainty, denoted as `Uncert.', accounts for the influence of the finite size of the basis set and is estimated as the difference between the values computed with the $3\xi$ and $4\xi$ basis sets. The values in the columns labeled `FINAL' and `Uncert.' are reported in Table III of the main text as the results of the e60-CCSD calculations for the electric dipole polarizabilities $\alpha_d^{S,T}$. \label{tab:Ir17+_alpha_DIRAC}}{\setlength{\tabcolsep}{3pt}
		\begin{tabular}{cccc cccc cccc cccc cc }\hline\hline  \addlinespace[0.1cm]
\multirow{2}{*}{Config.}	&	\multirow{2}{*}{Term}	&&	\multicolumn{3}{c}{$2\xi$}									&&	\multicolumn{3}{c}{$3\xi$}									&&	\multicolumn{3}{c}{$4\xi$ FINAL}									&&	\multicolumn{3}{c}{Uncert.}					\\ \cline{4-6}  \cline{8-10} \cline{12-14}  \cline{16-18}\addlinespace[0.1cm]
&		&&	$\alpha^S$	&	$\alpha^T$	&	$\Delta \alpha^S$	&&	$\alpha^S$	&	$\alpha^T$	&	$\Delta \alpha^S$	&&	$\alpha^S$	&	$\alpha^T$	&	$\Delta \alpha^S$	&&	$\alpha^S$	&	$\alpha^T$	&	$\Delta \alpha^S$	\\\hline \addlinespace[0.1cm]
$4f^{13}5s$	&	$^3F_4^o$	&&	0.3386 	&	0.0097 	&	0.0000 	&&	0.3713 	&	0.0013 	&	0.0000 	&&	0.3736 	&	0.0077 	&	0.0000 	&&	0.0024 	&	0.0064 	&	0.0000 	\\
&	$^3F_3^o$	&&	0.3399 	&	0.0054 	&	0.0013 	&&	0.3745 	&	-0.0077 	&	0.0032 	&&	0.3744 	&	0.0133 	&	0.0008 	&&	-0.0001 	&	0.0209 	&	-0.0024 	\\
&	$^3F_2^o$	&&	0.3400 	&	0.0054 	&	0.0014 	&&	0.3731 	&	0.0087 	&	0.0019 	&&	0.3731 	&	0.0161 	&	-0.0005 	&&	0.0000 	&	0.0074 	&	-0.0024 	\\
&	$^1F_3^o$	&&	0.3381 	&	0.0110 	&	-0.0006 	&&	0.3688 	&	-0.0069 	&	-0.0024 	&&	0.3747 	&	0.0080 	&	0.0011 	&&	0.0059 	&	0.0149 	&	0.0035 	\\ \addlinespace[0.1cm]
$4f^{14}$	&	$^1S_0$	&&	0.1104 	&	0.0000 	&	-0.2282 	&&	0.1615 	&	0.0000 	&	-0.2098 	&&	0.1734 	&	0.0000 	&	-0.2002 	&&	0.0119 	&	0.0000 	&	0.0096 	\\ \addlinespace[0.1cm]
$4f^{12}5s^2$	&	$^3H_6$	&&	0.4777 	&	0.0178 	&	0.1391 	&&	0.4952 	&	0.0109 	&	0.1239 	&&	0.4979 	&	0.0079 	&	0.1243 	&&	0.0027 	&	-0.0030 	&	0.0004 	\\
&	$^3F_4$	&&	0.4786 	&	0.0000 	&	0.1400 	&&	0.4990 	&	0.0643 	&	0.1277 	&&	0.4980 	&	0.0105 	&	0.1244 	&&	-0.0010 	&	-0.0538 	&	-0.0033 	\\
&	$^3H_5$	&&	0.4765 	&	0.0082 	&	0.1379 	&&	0.4955 	&	0.0106 	&	0.1242 	&&	0.4984 	&	0.0072 	&	0.1248 	&&	0.0030 	&	-0.0034 	&	0.0006 	\\
&	$^3F_2$	&&	0.4766 	&	-0.0174 	&	0.1380 	&&	0.4949 	&	0.0136 	&	0.1236 	&&	0.4976 	&	-0.0140 	&	0.1240 	&&	0.0027 	&	-0.0276 	&	0.0003 	\\
&	$^1G_4$	&&	0.4759 	&	0.0126 	&	0.1373 	&&	0.4973 	&	-0.0026 	&	0.1261 	&&	0.4984 	&	0.0108 	&	0.1248 	&&	0.0011 	&	0.0135 	&	-0.0013 	\\
&	$^3F_3$	&&	0.4765 	&	-0.0078 	&	0.1379 	&&	0.5001 	&	-0.0321 	&	0.1288 	&&	0.4989 	&	-0.0032 	&	0.1253 	&&	-0.0011 	&	0.0290 	&	-0.0035 	\\
&	$^3H_4$	&&	0.4758 	&	0.0181 	&	0.1372 	&&	0.4930 	&	-0.0071 	&	0.1217 	&&	0.4967 	&	0.0140 	&	0.1231 	&&	0.0037 	&	0.0211 	&	0.0014 	\\
&	$^1D_2$	&&	0.4771 	&	-0.0111 	&	0.1385 	&&	0.4987 	&	0.0083 	&	0.1275 	&&	0.4985 	&	-0.0086 	&	0.1249 	&&	-0.0002 	&	-0.0169 	&	-0.0026 	\\
&	$^1J_6$	&&	0.4772 	&	0.0202 	&	0.1386 	&&	0.4992 	&	0.0160 	&	0.1279 	&&	0.4982 	&	0.0161 	&	0.1245 	&&	-0.0010 	&	0.0001 	&	-0.0034 	\\ \hline \hline  
	\end{tabular}}
\end{table}

\subsection{$\Theta$}	
	
\begin{table}[H]
\centering	
\caption{The electric quadrupole moment $\Theta$ (in atomic units, a.u.) of the Ir$^{17+}$ ion is calculated using the KRCI method, based on the expectation value of the one-electron operator for the property. The e32-CISD calculations are performed with the dyall.aae2z ($2\xi$), dyall.aae3z ($3\xi$), and dyall.aae4z ($4\xi$) basis sets, considering truncations of virtual orbitals with energies less than 20 atomic units ($<20$) and 300 atomic units ($<300$). The values in the column corresponding to $4\xi<20$ are regarded as the final results, denoted as `FINAL.' The uncertainty, denoted as `Uncert.', is evaluated using the formula $\sqrt{\Delta_{virt}^2 + \Delta_{basis}^2}$. The parameter $\Delta_{virt}$ represents the influence of virtual-orbital truncation and is estimated as the difference between the values obtained under the conditions of $2\xi<20$ and $2\xi<300$. The parameter $\Delta_{basis}$ accounts for the effect of the finite size of the basis set and is estimated as the difference between the values obtained under the conditions of $3\xi<20$ and $4\xi<20$. The values in the columns labeled `FINAL' and `Uncert.' are reported in Table IV of the main text as the results of the e32-CISD calculations for the electric quadrupole moment $\Theta$.  \label{tab:Ir17+_theta_krci}}{\setlength{\tabcolsep}{12pt}
	    \begin{tabular}{ccc ccc ccc }\hline\hline  \addlinespace[0.1cm]
Config.	&	Term	&	$2\xi<20$	&	$2\xi<300$	&	$3\xi<20$	&	$4\xi<20$ FINAL 	&	$\Delta_{virt}$	&	$\Delta_{basis}$	&	Uncert.	\\ \hline \addlinespace[0.1cm]
$4f^{13}5s$	&	$^3F_4^o$	&	0.0915 	&	0.0912 	&	0.0906 	&	0.0871 	&	-0.0003 	&	-0.0035 	&	0.0036 	\\
&	$^3F_3^o$	&	0.0787 	&	0.0784 	&	0.0781 	&	0.0750 	&	-0.0003 	&	-0.0030 	&	0.0031 	\\
&	$^3F_2^o$	&	0.0616 	&	0.0613 	&	0.0610 	&	0.0586 	&	-0.0003 	&	-0.0024 	&	0.0025 	\\
&	$^1F_3^o$	&	0.0838 	&	0.0832 	&	0.0830 	&	0.0796 	&	-0.0006 	&	-0.0034 	&	0.0035 	\\
$4f^{14}$	&	$^1S_0$	&	0	&	0 	&	0 	&	0 	&	0 	&	0 	&	0 	\\ \addlinespace[0.1cm]
$4f^{12}5s^2$	&	$^3H_6$	&	0.0850 	&	0.0856 	&	0.0846 	&	0.0822 	&	0.0006 	&	-0.0023 	&	0.0024 	\\ \addlinespace[0.1cm]
&	$^3F_4$	&	-0.0096 	&	-0.0087 	&	-0.0095 	&	-0.0092 	&	0.0009 	&	0.0003 	&	0.0010 	\\
&	$^3H_5$	&	0.0747 	&	0.0752 	&	0.0743 	&	0.0722 	&	0.0005 	&	-0.0021 	&	0.0022 	\\
&	$^3F_2$	&	-0.0412 	&	-0.0401 	&	-0.0410 	&	-0.0398 	&	0.0011 	&	0.0012 	&	0.0016 	\\
&	$^1G_4$	&	0.0299 	&	0.0316 	&	0.0294 	&	0.0294 	&	0.0018 	&	0.0000 	&	0.0018 	\\
&	$^3F_3$	&	-0.0212 	&	-0.0205 	&	-0.0211 	&	-0.0201 	&	0.0006 	&	0.0010 	&	0.0012 	\\
&	$^3H_4$	&	0.0395 	&	0.0397 	&	0.0394 	&	0.0384 	&	0.0002 	&	-0.0010 	&	0.0010 	\\
&	$^1D_2$	&	-0.0269 	&	-0.0266 	&	-0.0268 	&	-0.0252 	&	0.0002 	&	0.0016 	&	0.0016 	\\
&	$^1J_6$	&	0.1623 	&	0.1633 	&	0.1614 	&	0.1565 	&	0.0010 	&	-0.0050 	&	0.0051 	\\\hline\hline
	\end{tabular}}
	\end{table}

\begin{table}[H]
\centering	
\caption{The electric quadrupole moment $\Theta$ (in atomic units, a.u.) of the Ir$^{17+}$ ion is obtained through e60-CCSD finite-field (FF) calculations. These calculations are performed using the DIRAC code with the dyall.aae2z ($2\xi$), dyall.aae3z ($3\xi$), and dyall.aae4z ($4\xi$) basis sets. The value obtained with the $4\xi$ basis set is regarded as the final value, denoted as `FINAL'. The uncertainty, labeled as `Uncert.', is estimated by considering the effect of the finite size of the basis set. Specifically, it is evaluated as the difference between the results obtained with the $3\xi$ and $4\xi$ basis sets. The values in the columns labeled `FINAL' and `Uncert.' are reported in Table IV of the main text as the results of the e60-CCSD calculations for the electric quadrupole moment $\Theta$.  \label{tab:Ir17+_theta_DIRAC}}{\setlength{\tabcolsep}{20pt}
	    \begin{tabular}{ccc ccc}\hline\hline  \addlinespace[0.1cm]
Config.	&	Term	&	$2\xi$	&	$3\xi$	&	$4\xi$ FINAL	&	Uncert.	\\ \hline \addlinespace[0.1cm]
$4f^{13}5s$	&	$^3F_4^o$	&	-0.0906 	&	-0.0874 	&	-0.0859 	&	0.0015 	\\
&	$^3F_3^o$	&	-0.0782 	&	-0.0756 	&	-0.0743 	&	0.0013 	\\
&	$^3F_2^o$	&	-0.0612 	&	-0.0590 	&	-0.0579 	&	0.0011 	\\
&	$^1F_3^o$	&	-0.0839 	&	-0.0808 	&	-0.0793 	&	0.0015 	\\ \addlinespace[0.1cm]
$4f^{14}$	&	$^1S_0$	&	0	&	0	&	0	&	0	\\ \addlinespace[0.1cm]
$4f^{12}5s^2$	&	$^3H_6$	&	-0.0841 	&	-0.0816 	&	-0.0804 	&	0.0012 	\\
&	$^3F_4$	&	0.0098 	&	0.0096 	&	0.0095 	&	-0.0002 	\\
&	$^3H_5$	&	-0.0741 	&	-0.0718 	&	-0.0707 	&	0.0011 	\\
&	$^3F_2$	&	0.0410 	&	0.0399 	&	0.0395 	&	-0.0005 	\\
&	$^1G_4$	&	-0.0296 	&	-0.0294 	&	-0.0290 	&	0.0004 	\\
&	$^3F_3$	&	0.0211 	&	0.0201 	&	0.0198 	&	-0.0003 	\\
&	$^3H_4$	&	-0.0392 	&	-0.0380 	&	-0.0375 	&	0.0005 	\\
&	$^1D_2$	&	0.0268 	&	0.0252 	&	0.0247 	&	-0.0005 	\\
&	$^1J_6$	&	-0.1611 	&	-0.1557 	&	-0.1533 	&	0.0024 	\\ \hline \hline  
	\end{tabular}}
	\end{table}

\begin{table}[H]
	\centering	
	\caption{The electric quadrupole moment $\Theta$ (a.u.) of the Ir$^{17+}$ ion is determined via e60-CCSD finite-field (FF) calculations. These calculations are performed using the EXPT code with the dyall.aae2z ($2\xi$), dyall.aae3z ($3\xi$), and dyall.aae4z ($4\xi$) basis sets. The result obtained with the $4\xi$ basis set is regarded as the final value, denoted as `FINAL'. The uncertainty, labeled as `Uncert.', accounts for the effect of the finite size of the basis set and is estimated from the difference between the results computed with the $3\xi$ and $4\xi$ basis sets.\label{tab:Ir17+_theta_EXPT}}{\setlength{\tabcolsep}{20pt}
		\begin{tabular}{ccc ccc}\hline\hline  \addlinespace[0.1cm]
Config.	&	Term	&	$2\xi$	&	$3\xi$	&	$4\xi$ FINAL	&	Uncert.	\\ \hline \addlinespace[0.1cm]
$4f^{13}5s$	&	$^3F_4^o$	&	-0.0901 	&	-0.0868 	&	-0.0853 	&	0.0015 	\\
&	$^3F_3^o$	&	-0.0780 	&	-0.0750 	&	-0.0738 	&	0.0012 	\\
&	$^3F_2^o$	&	-0.0608 	&	-0.0586 	&	-0.0576 	&	0.0010 	\\
&	$^1F_3^o$	&	-0.0833 	&	-0.0802 	&	-0.0788 	&	0.0014 	\\ \addlinespace[0.1cm]
$4f^{14}$	&	$^1S_0$	&	0	&	0	&	0	&0		\\ \addlinespace[0.1cm]
$4f^{12}5s^2$	&	$^3H_6$	&	-0.0836 	&	-0.0809 	&	-0.0798 	&	0.0011 	\\
&	$^3F_4$	&	0.0099 	&	0.0097 	&	0.0095 	&	-0.0002 	\\
&	$^3H_5$	&	-0.0736 	&	-0.0712 	&	-0.0702 	&	0.0010 	\\
&	$^3F_2$	&	0.0409 	&	0.0398 	&	0.0392 	&	-0.0006 	\\
&	$^1G_4$	&	-0.0292 	&	-0.0290 	&	-0.0291 	&	-0.0001 	\\
&	$^3F_3$	&	0.0211 	&	0.0201 	&	0.0197 	&	-0.0004 	\\
&	$^3H_4$	&	-0.0388 	&	-0.0376 	&	-0.0372 	&	0.0004 	\\
&	$^1D_2$	&	0.0261 	&	0.0252 	&	0.0246 	&	-0.0006 	\\
&	$^1J_6$	&	-0.1603 	&	-0.1547 	&	-0.1523 	&	0.0024 	\\ \hline \hline  
	\end{tabular}}
\end{table}

\subsection{$A$}	
	
\begin{table}[H]
\centering
\caption{The hyperfine structure constant $A$ (in megahertz, MHz) of the Ir$^{17+}$ ion is calculated using e32-CISD calculations within the KRCI method. These calculations are performed with the dyall.aae2z ($2\xi$), dyall.aae3z ($3\xi$), and dyall.aae4z ($4\xi$) basis sets, considering truncations of virtual orbitals with energies less than 20 a.u. ($<20$) and 300 a.u. ($<300$). The values corresponding to the $4\xi<20$ column are regarded as the final results and are denoted as `FINAL'. The uncertainty, labeled as `Uncert.', is estimated using the formula $\sqrt{\Delta_{virt}^2 + \Delta_{basis}^2}$. The data $\Delta_{virt}$ represents the effect of virtual-orbital truncation and is evaluated as the difference between the values obtained with $2\xi<20$ and $2\xi<300$. The data $\Delta_{basis}$ accounts for the influence of the finite size of the basis set and is estimated as the difference between the values obtained with $3\xi<20$ and $4\xi<20$. The values in the columns labeled `FINAL' and `Uncert.' are reported in Table IV of the main text as the results of the e32-CISD calculations for the hyperfine structure constant $A$.\label{tab:Ir17+_HFSA-KRCI}}{\setlength{\tabcolsep}{8pt}
	    \begin{tabular}{ccc ccc ccc c}\hline\hline  \addlinespace[0.1cm]
Config.	&	Term	&	$2\xi<20$ (e24)	&	$2\xi<20$	&	$2\xi<300$	&	$3\xi<20$	&	$4\xi<20$ FINAL	&	$\Delta_{virt}$	&	$\Delta_{basis}$	&	Uncert. 	\\ \hline \addlinespace[0.1cm]
$4f^{13}5s$	&	$^3F_4^o$	&	5241.5 	&	5154.5 	&	5300.9 	&	5184.3 	&	5192.8 	&	59.4 	&	8.5 	&	60.0 	\\
&	$^3F_3^o$	&	-4358.0 	&	-4269.3 	&	-4415.5 	&	-4306.1 	&	-4324.9 	&	-57.5 	&	-18.8 	&	60.5 	\\
&	$^3F_2^o$	&	-6281.0 	&	-6162.9 	&	-6361.4 	&	-6199.1 	&	-6209.7 	&	-80.4 	&	-10.6 	&	81.1 	\\
&	$^1F_3^o$	&	6613.8 	&	6492.2 	&	6692.2 	&	6542.0 	&	6564.5 	&	78.3 	&	22.6 	&	81.5 	\\ \addlinespace[0.1cm]
$4f^{14}$	&	$^1S_0$	&	0 	&	0 	&	0 	&	0 	&	0 	&	0 	&	0 	&	0 	\\ \addlinespace[0.1cm]
$4f^{12}5s^2$	&	$^3H_6$	&	246.8 	&	244.6 	&	244.9 	&	246.3 	&	245.3 	&	-1.8 	&	-1.1 	&	2.1 	\\
&	$^3F_4$	&	243.1 	&	241.0 	&	241.4 	&	242.6 	&	241.9 	&	-1.7 	&	-0.8 	&	1.8 	\\
&	$^3H_5$	&	294.4 	&	293.7 	&	294.1 	&	295.5 	&	295.6 	&	1.0 	&	0.1 	&	1.0 	\\
&	$^3F_2$	&	239.7 	&	238.5 	&	241.5 	&	240.4 	&	241.2 	&	1.8 	&	0.8 	&	1.9 	\\
&	$^1G_4$	&	291.2 	&	290.4 	&	291.3 	&	292.2 	&	292.5 	&	1.0 	&	0.3 	&	1.0 	\\
&	$^3F_3$	&	271.1 	&	269.7 	&	270.2 	&	271.5 	&	271.1 	&	-1.0 	&	-0.4 	&	1.0 	\\
&	$^3H_4$	&	374.1 	&	376.3 	&	375.1 	&	378.3 	&	379.3 	&	1.0 	&	1.0 	&	1.4 	\\
&	$^1D_2$	&	271.1 	&	269.0 	&	269.5 	&	270.8 	&	270.1 	&	-1.6 	&	-0.7 	&	1.7 	\\
&	$^1J_6$	&	296.9 	&	296.5 	&	296.5 	&	298.3 	&	298.4 	&	1.0 	&	0.1 	&	1.0 	\\ \hline \hline  
	\end{tabular}}
	\end{table}

\begin{table}[H]
\centering
\caption{The hyperfine structure constant $A$ (in MHz) of the Ir$^{17+}$ ion is derived from e60-CCSD finite-field (FF) calculations. These calculations are performed using the DIRAC code with the dyall.aae3z ($3\xi$) and dyall.aae4z ($4\xi$) basis sets. The value obtained with the $4\xi$ basis set is regarded as the final result, denoted as `FINAL'. The uncertainty, labeled as `Uncert.', accounts for the effect of the finite size of the basis set and is estimated as the difference between the results computed with the $3\xi$ and $4\xi$ basis sets. The values in the columns labeled `FINAL' and `Uncert.' are reported in Table IV of the main text as the results of the e60-CCSD calculations for the hyperfine structure constant $A$. \label{tab:Ir17+_HFSA_DIRAC}}{\setlength{\tabcolsep}{20pt}
		\begin{tabular}{ccc cc}\hline\hline  \addlinespace[0.1cm]
Config.	&	Term	&	$3\xi$	&	$4\xi$ FINAL	&	Uncert.	\\ \hline \addlinespace[0.1cm]	
$4f^{13}5s$	&	$^3F_4^o$	&	5270.5 	&	5268.4 	&	-2.1 	\\	
&	$^3F_3^o$	&	-4421.5 	&	-4434.4 	&	-12.9 	\\	
&	$^3F_2^o$	&	-6350.3 	&	-6346.5 	&	3.8 	\\	
&	$^1F_3^o$	&	6665.6 	&	6676.2 	&	10.6 	\\ \addlinespace[0.1cm]	
$4f^{14}$	&	$^1S_0$	&	0 	&	0 	&	0	\\ \addlinespace[0.1cm]	
$4f^{12}5s^2$	&	$^3H_6$	&	221.2 	&	219.7 	&	-1.5 	\\	
&	$^3F_4$	&	220.7 	&	219.3 	&	-1.4 	\\	
&	$^3H_5$	&	278.8 	&	279.1 	&	0.3 	\\	
&	$^3F_2$	&	254.7 	&	253.9 	&	-0.8 	\\	
&	$^1G_4$	&	281.1 	&	281.4 	&	0.3 	\\	
&	$^3F_3$	&	252.3 	&	251.9 	&	-0.4 	\\	
&	$^3H_4$	&	363.0 	&	365.9 	&	2.9 	\\	
&	$^1D_2$	&	245.2 	&	244.1 	&	-1.1 	\\	
&	$^1J_6$	&	284.9 	&	285.5 	&	0.6 	\\ \hline \hline  	
	\end{tabular}}
\end{table}

\begin{table}[H]
\centering	
\caption{The field shift ($F$, in GHz) of the Ir$^{17+}$ ion is calculated using e60-CCSD finite-field (FF) calculations. These calculations are performed with the DIRAC code using the dyall.aae2z ($2\xi$), dyall.aae3z ($3\xi$), and dyall.aae4z ($4\xi$) basis sets. The value obtained with the $4\xi$ basis set is regarded as the final result, denoted as `FINAL'. The uncertainty, labeled as `Uncert.', accounts for the effect of the finite size of the basis set and is estimated as the difference between the results obtained with the $3\xi$ and $4\xi$ basis sets. The values in the columns labeled `FINAL' and `Uncert.' are reported in Table IV of the main text as the results of the e60-CCSD calculations for the field shift $F$.\label{tab:Ir17+_FS_DIRAC}}{\setlength{\tabcolsep}{20pt}
\begin{tabular}{ccc ccc}\hline\hline  \addlinespace[0.1cm]
Config.	&	Term	&	$2\xi$	&	$3\xi$	&	$4\xi$-FINAL	&	Uncert.	\\ \hline \addlinespace[0.1cm]
$4f^{13}5s$	&	$^3F_4^o$	&	0.00 	&	0.00 	&	0.00 	&	0.00 	\\
&	$^3F_3^o$	&	-1.81 	&	-1.92 	&	-1.92 	&	0.00 	\\
&	$^3F_2^o$	&	1.47 	&	1.45 	&	1.47 	&	0.02 	\\
&	$^1F_3^o$	&	-1.92 	&	-1.94 	&	-1.90 	&	0.04 	\\ \addlinespace[0.1cm]
$4f^{14}$	&	$^1S_0$	&	-614.90 	&	-617.66 	&	-618.09 	&	-0.43 	\\ \addlinespace[0.1cm]
$4f^{12}5s^2$	&	$^3H_6$	&	639.05 	&	643.39 	&	643.64 	&	0.25 	\\
&	$^3F_4$	&	639.12 	&	643.48 	&	643.73 	&	0.25 	\\
&	$^3H_5$	&	640.47 	&	644.83 	&	645.06 	&	0.23 	\\
&	$^3F_2$	&	639.73 	&	644.11 	&	644.36 	&	0.25 	\\
&	$^1G_4$	&	640.65 	&	645.06 	&	645.28 	&	0.23 	\\
&	$^3F_3$	&	640.59 	&	644.99 	&	645.22 	&	0.23 	\\
&	$^3H_4$	&	642.01 	&	646.41 	&	646.66 	&	0.25 	\\
&	$^1D_2$	&	640.88 	&	645.26 	&	645.51 	&	0.25 	\\
&	$^1J_6$	&	641.10 	&	645.47 	&	645.71 	&	0.25 	\\ \hline \hline  
	\end{tabular}}
\end{table}

\begin{table}[H]
	\centering	
	\caption{The field shift ($F$, in GHz) of the Ir$^{17+}$ ion is obtained using the e60-CCSD finite-field (FF) calculation, implemented with the EXPT code and the dyall.aae2z ($2\xi$) and dyall.aae3z ($3\xi$) basis sets.  \label{tab:Ir17+_FS_EXPT}}{\setlength{\tabcolsep}{20pt}
		\begin{tabular}{ccc cc}\hline\hline  \addlinespace[0.1cm]
Config.	&	Term	&	$2\xi$	&	$3\xi$	\\ \hline \addlinespace[0.1cm]
$4f^{13}5s$	&	$^3F_4^o$	&	0.0 	&	0.0 	\\
&	$^3F_3^o$	&	-1.8 	&	-1.9 	\\
&	$^3F_2^o$	&	1.4 	&	1.4 	\\
&	$^1F_3^o$	&	-1.9 	&	-2.0 	\\ \addlinespace[0.1cm]
$4f^{14}$	&	$^1S_0$	&	-614.9 	&	-617.7 	\\ \addlinespace[0.1cm]
$4f^{12}5s^2$	&	$^3H_6$	&	639.0 	&	643.4 	\\
&	$^3F_4$	&	640.4 	&	643.5 	\\
&	$^3H_5$	&	639.7 	&	644.8 	\\
&	$^3F_2$	&	640.6 	&	644.1 	\\
&	$^1G_4$	&	640.5 	&	645.0 	\\
&	$^3F_3$	&	641.9 	&	645.0 	\\
&	$^3H_4$	&	640.8 	&	646.4 	\\
&	$^1D_2$	&	641.1 	&	645.3 	\\
&	$^1J_6$	&	641.1 	&	645.5 	\\ \hline \hline  
	\end{tabular}}
\end{table}

\newpage
\section{Calculation of lifetime of excited states}	
In this section, we provide detailed data to estimate the lifetime $\tau$ of the excited states of the Ir$^{17+}$ ion. Using the e32-CISD $<20$ approach within the KRCI method, in combination with the $4\xi$ basis set, we calculate the electric dipole (E1), magnetic dipole (M1), and electric quadrupole (E2) transition matrix elements. In our study, the uncertainty in the lifetime $\tau$ primarily stems from the uncertainty in the EE. Therefore, we evaluate the uncertainty in $\tau$ based on the uncertainty in the EE. The corresponding EE values are provided in Table \ref{tab:Ir17+_EE_KRCI}.

The lifetime $\tau$ (in seconds) of an excited state within the Ir$^{17+}$ configuration is computed using the following formula:
\begin{eqnarray}
\tau = \frac{1}{\sum_{i,O} A_{fi}^O},
\label{eq:tau}
\end{eqnarray}
where $\lambda_{fi}$ (in angstroms) represents the transition wavelength between the upper state $f$ and the lower state $i$, and $A_{fi}^O$ denotes the transition probability for channel $O$, given by the following expressions:
\begin{eqnarray}
A_{fi}^{M1} = \frac{2.69735 \times 10^{13}}{(2J_f + 1)\lambda_{fi}^3} S_{fi}^{M1},
\label{AM1}
\end{eqnarray}
\begin{eqnarray}
A_{fi}^{E1} = \frac{2.02613 \times 10^{18}}{(2J_f + 1)\lambda_{fi}^3} S_{fi}^{E1},
\label{AE1}
\end{eqnarray}
and
\begin{eqnarray}
A_{fi}^{E2} = \frac{1.11995 \times 10^{18}}{(2J_f + 1)\lambda_{fi}^5} S_{fi}^{E2},
\label{AE2}
\end{eqnarray}
where the line strengths for the M1 and E2 transitions are defined as $S_{fi}^{M1} = \langle M1 \rangle^2$ and $S_{fi}^{E2} = \langle E2 \rangle^2$, respectively.

\setlength{\tabcolsep}{8pt} 
\setlength{\LTcapwidth}{\textwidth}
\setlength{\LTpre}{0pt}
\setlength{\LTpost}{0pt}
\begin{longtable}{llcccccc}
\caption{The wavelengths ($\lambda$, in \(\mathring{\mathrm{A}}\)), transition rates ($A_r$, in s$^{-1}$), energies of the lower and upper levels (in cm$^{-1}$), lifetimes ($\tau$, in s), and branching ratios for the M1, E1, and E2 transitions of the Ir$^{17+}$ ion are presented. In the table, the symbol `–’ indicates that no physically allowed transition channels exist between the two states for the specified transition type, due to selection rules or other factors.\label{tab:Ir17+_tau}}\\ \hline \hline \addlinespace[0.1cm]
up state ($f$)	&	down state $(i)$           	&	    $\lambda$ (\AA) 	&	    $A_{fi}^{E1}$   (/s) 	&	    $A_{fi}^{M1}$   (/s) 	&	$A_{fi}^{E2}$   (/s)	&	$A_{fi}^{T}$  (/s) 	&	$\tau_f$ (s)	\\ \hline \addlinespace[0.1cm]
\endfirsthead

\multicolumn{8}{c}%
{{\bfseries \tablename\ \thetable{} -- continued from previous page}} \\ \hline \addlinespace[0.1cm]
up state ($f$)	&	down state $(i)$           	&	    $\lambda$ (\AA) 	&	    $A_{fi}^{E1}$   (/s) 	&	    $A_{fi}^{M1}$   (/s) 	&	$A_{fi}^{E2}$   (/s)	&	$A_{fi}^{T}$  (/s) 	&	$\tau_f$ (s)	\\ \hline \addlinespace[0.1cm]
\endhead

\hline \multicolumn{8}{r}{{Continued on next page}} \\ \hline
\endfoot

\hline \hline
\endlastfoot

G1: $(4f^{13}5s) ^3F_3^o$	&	G0: $(4f^{13}5s) ^3F_4^o$	&	2.14E+04	&	--	&	1.00E+00	&	--	&	1.00E+00	&	6.10E-01	\\[+2ex]
G2: $(4f^{13}5s) ^3F_2^o$	&	G0: $(4f^{13}5s) ^3F_4^o$	&	3.83E+03	&	--	&	--	&	--	&	--	&	4.60E-03	\\
	&	G1: $(4f^{13}5s) ^3F_3^o$	&	4.66E+03	&	--	&	1.00E+00	&	--	&	1.00E+00	&		\\
	&	E1: $(4f^{14}) ^1S_0$	&	9.02E+03	&	--	&	--	&	--	&	--	&		\\[+2ex]
G3: $(4f^{13}5s) ^1F_3^o$	&	G0: $(4f^{13}5s) ^3F_4^o$	&	3.20E+03	&	--	&	9.64E-01	&	--	&	9.64E-01	&	3.24E-03	\\
	&	G1: $(4f^{13}5s) ^3F_3^o$	&	3.77E+03	&	--	&	3.18E-02	&	--	&	3.18E-02	&		\\
	&	G2: $(4f^{13}5s) ^3F_2^o$	&	1.97E+04	&	--	&	4.10E-03	&	--	&	4.10E-03	&		\\
	&	E1: $(4f^{14}) ^1S_0$	&	6.19E+03	&	--	&	--	&	--	&	--	&		\\
	&	E2: $(4f^{12}5s^2) ^3H_6$	&	2.48E+04	&	--	&	--	&	--	&	--	&		\\[+2ex]
E1: $(4f^{14}) ^1S_0$	&	G0: $(4f^{13}5s) ^3F_4^o$	&	6.64E+03	&	--	&	--	&	--	&	--	&	Infinity	\\
	&	G1: $(4f^{13}5s) ^3F_3^o$	&	9.64E+03	&	--	&	--	&	--	&	--	&		\\[+2ex]
E2: $(4f^{12}5s^2) ^3H_6$	&	G0: $(4f^{13}5s) ^3F_4^o$	&	3.68E+03	&	--	&	--	&	--	&	--	&	Infinity	\\
	&	G1: $(4f^{13}5s) ^3F_3^o$	&	4.45E+03	&	--	&	--	&	--	&	--	&		\\
	&	G2: $(4f^{13}5s) ^3F_2^o$	&	9.69E+04	&	--	&	--	&	--	&	--	&		\\
	&	E1: $(4f^{14}) ^1S_0$	&	8.25E+03	&	--	&	--	&	--	&	--	&		\\[+2ex]
E3: $(4f^{12}5s^2) ^3F_4$	&	G0: $(4f^{13}5s) ^3F_4^o$	&	2.71E+03	&	5.44E-01	&	--	&	--	&	5.44E-01	&	1.04E-01	\\
	&	G1: $(4f^{13}5s) ^3F_3^o$	&	3.10E+03	&	4.56E-01	&	--	&	--	&	4.56E-01	&		\\
	&	G2: $(4f^{13}5s) ^3F_2^o$	&	9.28E+03	&	--	&	--	&	--	&	--	&		\\
	&	G3: $(4f^{13}5s) ^1F_3^o$	&	1.75E+04	&	4.00E-04	&	--	&	--	&	4.00E-04	&		\\
	&	E1: $(4f^{14}) ^1S_0$	&	4.57E+03	&	--	&	--	&	--	&	--	&		\\
	&	E2: $(4f^{12}5s^2) ^3H_6$	&	1.03E+04	&	--	&	--	&	--	&	--	&		\\[+2ex]
E4: $(4f^{12}5s^2) ^3H_5$	&	G0: $(4f^{13}5s) ^3F_4^o$	&	1.94E+03	&	4.00E-04	&	--	&	--	&	4.00E-04	&	2.66E-03	\\
	&	G1: $(4f^{13}5s) ^3F_3^o$	&	2.14E+03	&	--	&	--	&	--	&	--	&		\\
	&	G2: $(4f^{13}5s) ^3F_2^o$	&	3.94E+03	&	--	&	--	&	--	&	--	&		\\
	&	G3: $(4f^{13}5s) ^1F_3^o$	&	4.93E+03	&	--	&	--	&	--	&	--	&		\\
	&	E1: $(4f^{14}) ^1S_0$	&	2.74E+03	&	--	&	--	&	--	&	--	&		\\
	&	E2: $(4f^{12}5s^2) ^3H_6$	&	4.11E+03	&	--	&	9.87E-01	&	--	&	9.87E-01	&		\\
	&	E3: $(4f^{12}5s^2) ^3F_4$	&	6.86E+03	&	--	&	1.25E-02	&	--	&	1.25E-02	&		\\[+2ex]
E5: $(4f^{12}5s^2) ^3F_2$	&	G0: $(4f^{13}5s) ^3F_4^o$	&	1.70E+03	&	--	&	--	&	--	&	--	&	4.88E-02	\\
	&	G1: $(4f^{13}5s) ^3F_3^o$	&	1.85E+03	&	5.73E-01	&	--	&	--	&	5.73E-01	&		\\
	&	G2: $(4f^{13}5s) ^3F_2^o$	&	3.06E+03	&	3.55E-01	&	--	&	--	&	3.55E-01	&		\\
	&	G3: $(4f^{13}5s) ^1F_3^o$	&	3.62E+03	&	7.20E-02	&	--	&	--	&	7.20E-02	&		\\
	&	E1: $(4f^{14}) ^1S_0$	&	2.29E+03	&	--	&	--	&	--	&	--	&		\\
	&	E2: $(4f^{12}5s^2) ^3H_6$	&	3.16E+03	&	--	&	--	&	--	&	--	&		\\
	&	E3: $(4f^{12}5s^2) ^3F_4$	&	4.57E+03	&	--	&	--	&	3.00E-04	&	3.00E-04	&		\\
	&	E4: $(4f^{12}5s^2) ^3H_5$	&	1.37E+04	&	--	&	--	&	--	&	--	&		\\[+2ex]
E6: $(4f^{12}5s^2) ^1G_4$	&	G0: $(4f^{13}5s) ^3F_4^o$	&	1.65E+03	&	4.16E-01	&	--	&	--	&	4.16E-01	&	1.82E-03	\\
	&	G1: $(4f^{13}5s) ^3F_3^o$	&	1.79E+03	&	2.99E-01	&	--	&	--	&	2.99E-01	&		\\
	&	G2: $(4f^{13}5s) ^3F_2^o$	&	2.91E+03	&	--	&	--	&	--	&	--	&		\\
	&	G3: $(4f^{13}5s) ^1F_3^o$	&	3.41E+03	&	7.30E-03	&	--	&	--	&	7.30E-03	&		\\
	&	E1: $(4f^{14}) ^1S_0$	&	2.20E+03	&	--	&	--	&	--	&	--	&		\\
	&	E2: $(4f^{12}5s^2) ^3H_6$	&	3.00E+03	&	--	&	--	&	--	&	--	&		\\
	&	E3: $(4f^{12}5s^2) ^3F_4$	&	4.23E+03	&	--	&	2.59E-01	&	--	&	2.59E-01	&		\\
	&	E4: $(4f^{12}5s^2) ^3H_5$	&	1.11E+04	&	--	&	1.86E-02	&	--	&	1.88E-02	&		\\
	&	E5: $(4f^{12}5s^2) ^3F_2$	&	5.83E+04	&	--	&	--	&	--	&	--	&		\\[+2ex]
E7: $(4f^{12}5s^2) ^3F_3$	&	G0: $(4f^{13}5s) ^3F_4^o$	&	1.59E+03	&	3.17E-01	&	--	&	--	&	3.17E-01	&	1.58E-03	\\
	&	G1: $(4f^{13}5s) ^3F_3^o$	&	1.72E+03	&	2.34E-01	&	--	&	--	&	2.34E-01	&		\\
	&	G2: $(4f^{13}5s) ^3F_2^o$	&	2.72E+03	&	8.00E-04	&	--	&	--	&	8.00E-04	&		\\
	&	G3: $(4f^{13}5s) ^1F_3^o$	&	3.16E+03	&	1.93E-02	&	--	&	--	&	1.93E-02	&		\\
	&	E1: $(4f^{14}) ^1S_0$	&	2.09E+03	&	--	&	--	&	--	&	--	&		\\
	&	E2: $(4f^{12}5s^2) ^3H_6$	&	2.80E+03	&	--	&	--	&	--	&	--	&		\\
	&	E3: $(4f^{12}5s^2) ^3F_4$	&	3.85E+03	&	--	&	4.27E-01	&	--	&	4.27E-01	&		\\
	&	E4: $(4f^{12}5s^2) ^3H_5$	&	8.78E+03	&	--	&	--	&	--	&	--	&		\\
	&	E5: $(4f^{12}5s^2) ^3F_2$	&	2.46E+04	&	--	&	1.50E-03	&	--	&	1.50E-03	&		\\
	&	E6: $(4f^{12}5s^2) ^1G_4$	&	4.26E+04	&	--	&	2.00E-04	&	--	&	2.00E-04	&		\\[+2ex]
E8: $(4f^{12}5s^2) ^3H_4$	&	G0: $(4f^{13}5s) ^3F_4^o$	&	1.19E+03	&	2.40E-01	&	--	&	--	&	2.40E-01	&	5.37E-04	\\
	&	G1: $(4f^{13}5s) ^3F_3^o$	&	1.26E+03	&	3.48E-01	&	--	&	--	&	3.48E-01	&		\\
	&	G2: $(4f^{13}5s) ^3F_2^o$	&	1.73E+03	&	--	&	--	&	--	&	--	&		\\
	&	G3: $(4f^{13}5s) ^1F_3^o$	&	1.90E+03	&	2.57E-02	&	--	&	--	&	2.57E-02	&		\\
	&	E1: $(4f^{14}) ^1S_0$	&	1.45E+03	&	--	&	--	&	--	&	--	&		\\
	&	E2: $(4f^{12}5s^2) ^3H_6$	&	1.76E+03	&	--	&	--	&	--	&	--	&		\\
	&	E3: $(4f^{12}5s^2) ^3F_4$	&	2.12E+03	&	--	&	9.20E-03	&	--	&	9.20E-03	&		\\
	&	E4: $(4f^{12}5s^2) ^3H_5$	&	3.08E+03	&	--	&	2.98E-01	&	--	&	3.05E-01	&		\\
	&	E5: $(4f^{12}5s^2) ^3F_2$	&	3.97E+03	&	--	&	--	&	--	&	--	&		\\
	&	E6: $(4f^{12}5s^2) ^1G_4$	&	4.26E+03	&	--	&	6.46E-02	&	--	&	6.46E-02	&		\\
	&	E7: $(4f^{12}5s^2) ^3F_3$	&	4.74E+03	&	--	&	7.30E-03	&	--	&	7.30E-03	&		\\[+2ex]
E9: $(4f^{12}5s^2) ^1D_2$	&	G0: $(4f^{13}5s) ^3F_4^o$	&	1.15E+03	&	--	&	--	&	--	&	--	&	1.53E-03	\\
	&	G1: $(4f^{13}5s) ^3F_3^o$	&	1.22E+03	&	6.12E-02	&	--	&	--	&	6.12E-02	&		\\
	&	G2: $(4f^{13}5s) ^3F_2^o$	&	1.64E+03	&	1.64E-01	&	--	&	--	&	1.64E-01	&		\\
	&	G3: $(4f^{13}5s) ^1F_3^o$	&	1.79E+03	&	2.27E-01	&	--	&	--	&	2.27E-01	&		\\
	&	E1: $(4f^{14}) ^1S_0$	&	1.39E+03	&	--	&	--	&	--	&	--	&		\\
	&	E2: $(4f^{12}5s^2) ^3H_6$	&	1.67E+03	&	--	&	--	&	--	&	--	&		\\
	&	E3: $(4f^{12}5s^2) ^3F_4$	&	2.00E+03	&	--	&	--	&	6.00E-04	&	6.00E-04	&		\\
	&	E4: $(4f^{12}5s^2) ^3H_5$	&	2.82E+03	&	--	&	--	&	--	&	--	&		\\
	&	E5: $(4f^{12}5s^2) ^3F_2$	&	3.55E+03	&	--	&	2.95E-01	&	--	&	2.95E-01	&		\\
	&	E6: $(4f^{12}5s^2) ^1G_4$	&	3.79E+03	&	--	&	--	&	--	&	--	&		\\
	&	E7: $(4f^{12}5s^2) ^3F_3$	&	4.15E+03	&	--	&	2.53E-01	&	--	&	2.53E-01	&		\\
	&	E8: $(4f^{12}5s^2) ^3H_4$	&	3.38E+04	&	--	&	--	&	--	&	--	&		\\[+2ex]
E10: $(4f^{12}5s^2) ^1J_6$	&	G0: $(4f^{13}5s) ^3F_4^o$	&	1.03E+03	&	--	&	--	&	--	&	--	&	4.58E-03	\\
	&	G1: $(4f^{13}5s) ^3F_3^o$	&	1.08E+03	&	--	&	--	&	--	&	--	&		\\
	&	G2: $(4f^{13}5s) ^3F_2^o$	&	1.41E+03	&	--	&	--	&	--	&	--	&		\\
	&	G3: $(4f^{13}5s) ^1F_3^o$	&	1.52E+03	&	--	&	--	&	--	&	--	&		\\
	&	E1: $(4f^{14}) ^1S_0$	&	1.22E+03	&	--	&	--	&	--	&	--	&		\\
	&	E2: $(4f^{12}5s^2) ^3H_6$	&	1.43E+03	&	--	&	8.29E-01	&	2.00E-04	&	8.29E-01	&		\\
	&	E3: $(4f^{12}5s^2) ^3F_4$	&	1.66E+03	&	--	&	--	&	2.00E-04	&	2.00E-04	&		\\
	&	E4: $(4f^{12}5s^2) ^3H_5$	&	2.19E+03	&	1.80E-03	&	1.69E-01	&	--	&	1.71E-01	&		\\
	&	E5: $(4f^{12}5s^2) ^3F_2$	&	2.61E+03	&	--	&	--	&	--	&	--	&		\\
	&	E6: $(4f^{12}5s^2) ^1G_4$	&	2.73E+03	&	--	&	--	&	--	&	--	&		\\
	&	E7: $(4f^{12}5s^2) ^3F_3$	&	2.92E+03	&	--	&	--	&	--	&	--	&		\\
	&	E8: $(4f^{12}5s^2) ^3H_4$	&	7.58E+03	&	--	&	--	&	--	&	--	&		\\
	&	E9: $(4f^{12}5s^2) ^1D_2$	&	9.78E+03	&	--	&	--	&	--	&	--	&		\\ \hline \hline
\end{longtable}